\documentclass[manuscript]{acmart}
\usepackage{booktabs}
\usepackage{graphicx}
\AtBeginDocument{%
  }

\copyrightyear{2026}
\acmYear{2026}
\setcopyright{cc}
\setcctype{by}
\acmConference[CHI '26]{Proceedings of the 2026 CHI Conference on Human Factors in Computing Systems}{April 13--17, 2026}{Barcelona, Spain}
\acmBooktitle{Proceedings of the 2026 CHI Conference on Human Factors in Computing Systems (CHI '26), April 13--17, 2026, Barcelona, Spain}
\acmDOI{10.1145/3772318.3790972}
\acmISBN{979-8-4007-2278-3/2026/04}

\makeatletter
\renewenvironment{quote}
  {\list{}{\rightmargin=0pt
           \leftmargin=1.5em}%
   \item\relax}
  {\endlist}
\makeatother

\begin{document}

\title[Small Talk, Big Impact?]{Small Talk, Big Impact? LLM-based Conversational Agents to Mitigate Passive Fatigue in Conditional Automated Driving}


\author{Lewis Cockram}
\authornote{Both authors contributed equally to this research}
\orcid{0009-0006-4014-6196}
\affiliation{
  \institution{Queensland University of Technology}
  \city{Brisbane}
  \country{Australia}
}

\author{Yueteng Yu}
\authornotemark[1]
\orcid{0009-0001-3213-3276}
\affiliation{%
    \institution{Centre for Future Mobility}
    \institution{Queensland University of Technology}
    \city{Brisbane}
    \country{Australia}
}

\author{Jorge Pardo}
\orcid{0000-0001-7621-1005}
\affiliation{
  \institution{Centre for Future Mobility}
  \institution{Queensland University of Technology}
  \city{Brisbane}
  \country{Australia}
}

\author{Xiaomeng Li}
\orcid{0000-0003-2129-1671}
\affiliation{%
    \institution{Centre for Future Mobility}
    \institution{Queensland University of Technology}
    \city{Brisbane}
    \country{Australia}
}

\author{Andry Rakotonirainy}
\orcid{0000-0002-2144-4909}
\affiliation{%
    \institution{Centre for Future Mobility}
    \institution{Queensland University of Technology}
    \city{Brisbane}
    \country{Australia}
}

\author{Jonny Kuo}
\orcid{0009-0001-6422-4623}
\affiliation{%
    \institution{Seeing Machines Ltd.}
    \city{Melbourne}
    \country{Australia}
}

\author{Sébastien Demmel}
\orcid{0000-0001-8626-5306}
\affiliation{%
    \institution{Centre for Future Mobility}
    \institution{Queensland University of Technology}
    \city{Brisbane}
    \country{Australia}
}

\author{Michael G. Lenné}
\orcid{0000-0003-1671-6276}
\affiliation{%
    \institution{Seeing Machines Ltd.}
    \city{Melbourne}
    \country{Australia}
}

\author{Ronald Schroeter}
\email{r.schroeter@qut.edu.au}
\orcid{0000-0001-7990-1474}
\affiliation{%
    \institution{Centre for Future Mobility}
    \institution{Queensland University of Technology}
    \city{Brisbane}
    \country{Australia}
}
\renewcommand{\shortauthors}{Cockram \& Yu et al.}

\begin{abstract}
Passive fatigue during conditional automated driving can compromise driver readiness and safety. This paper presents findings from a test-track study with 40 participants in a real-world automated driving scenario. In this scenario, a Large Language Model (LLM) based conversational agent (CA) was designed to check in with drivers and re-engage them with their surroundings. Drawing on in-car video recordings, sleepiness ratings and interviews, we analysed how drivers interacted with the agent and how these interactions shaped alertness. Results show the CA is helpful for supporting vigilance during passive fatigue. Thematic analysis of acceptability further revealed three user preference profiles that implicate future intention to use CAs. Positioning empirically observed profiles within existing CA archetype frameworks highlights the need for adaptive design sensitive to diverse user groups. This work underscores the potential of CAs as proactive Human–Machine Interface (HMI) interventions, demonstrating how natural language can support context-aware interaction during automated driving.
\end{abstract}



\begin{teaserfigure}
    \centering
  \includegraphics[width=0.96\textwidth]{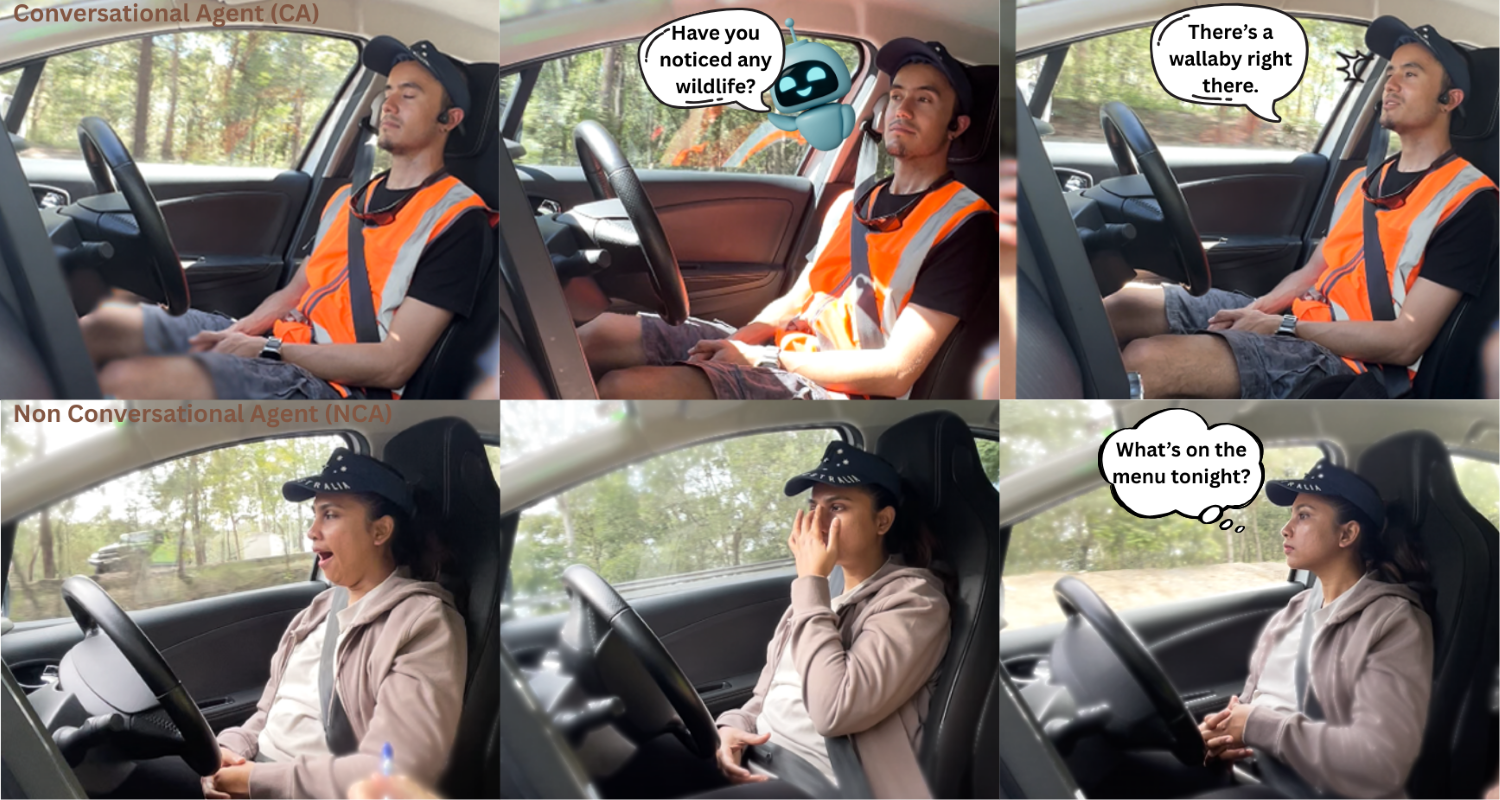}
  \caption{Using a conversational agent to reduce passive fatigue in a L3 automated driving scenario.}
  \Description{Two rows of images showing participants seated in an automated vehicle. The top row depicts a driver interacting with a conversational agent, with speech bubbles indicating questions about wildlife and comments about a wallaby. The bottom row shows a driver in the non-conversational condition, with speech bubbles reflecting non-driving internal thoughts. The contrast illustrates how the presence or absence of conversational prompts shapes engagement.}
  \label{fig:teaser}
\end{teaserfigure}

\maketitle

\section{Introduction}

While automated vehicles (AVs) can reduce human error-related crashes, a paradox is introduced. By relieving drivers of manual control and reducing human error, SAE Level 3 (L3) \cite{SAEJ3016_202104} \textit{Conditional Driving Automation} creates conditions for \textit{passive fatigue} \cite{Vogelpohl2019-ap}. This state of cognitive underload, monotony, and vigilance decline undermines drivers' readiness to take over \cite{May2009-vc}. Addressing passive fatigue is not simply a technical challenge but a safety-critical issue in Human-AV interactions \cite{Hancock2017-fv, Vogelpohl2019-ap}.

Various countermeasures for passive fatigue have been proposed, such as non-driving-related tasks (NDRTs) \cite{Weinbeer2019-gy,Loew2024-ew,Naujoks2018-su,large2016stimulating} or gamification \cite{Schroeter2016} to stimulate driver engagement during automated driving. These can temporarily mitigate cognitive underload in automated driving \cite{McKerral2023}. Additionally, few can adapt to the context or needs of individual drivers \cite{Gang2018-kt}. This challenge can be conceptualised using \emph{Malleable Attentional Resource Theory} (MART), which posits that attentional resources expand and contract in response to task demand \cite{Young2002}. Under low-demand conditions typical of automation, attentional resources contract, degrading vigilance and situational awareness, whereas overly demanding tasks can increase workload and induce active fatigue \cite{May2009-vc,Merat2019-ph}. Addressing the risks of underload situations is therefore as critical as mitigating those arising from excessive workload during control transitions. As a result, alternative forms of \textit{implicit} intervention for AVs have gained attention. Leveraging natural response or social stimulation, implicit interventions can deliver subtle 'nudges' that guide users to maintain awareness unobtrusively \cite{Stampf2022-xm}.

\textit{Conversational agents} (CAs) offer a novel and dynamic approach to delivering these nudges. CAs may engage users through natural social interaction with real-time, context-aware information about the environment\footnote{https://openai.com/index/hello-gpt-4o/}. Their design can draw on established archetypes, service-oriented, entertainment-focused, or socially companionable, each offering distinct ways to sustain engagement \cite{Lessio2020-xn}. 

Despite CAs' potential, previous studies have largely relied on driving simulators or Wizard-of-Oz (WoZ) setups. This is due to limited access to AV-capable vehicles and the technical challenges of deploying real-time AI interaction \cite{Large2019-ou, Huang2024-qe}. However, in intervention studies, simulators may suppress behavioural responses due to the absence of real motion and perceived risk, while WoZ setups constrain experimental control, making it difficult to replicate the precise timing and behavioural nuances of automated systems. Thus, the potential of real-time, large language model (LLM)-based agents in L3 automated driving remains unexplored \cite{Braun2019-oh}. This leaves unanswered questions about their effectiveness, usability, and acceptance in safety-critical contexts; a gap this paper aims to address. 

Our study explores the following research questions (RQs):

\begin{itemize}
    \item RQ1. How do users perceive the usage of a CA for supporting attention and engagement during monotonous automated driving?
    \item RQ2. What design and interaction factors shape users’ perceptions of using a CA in the context of monotonous automated driving?
    \item RQ3. What effects do safety-oriented CA interventions have on drivers’ behaviour and the states of alertness during authentic automated driving?
\end{itemize}

In this study, we introduced a real-time CA designed to maintain cognitive engagement during \textit{automated driving}. 40 participants interacted with it during a monotonous, driving task on a closed test track. The voice-based conversations were short, contextually grounded, and aimed at supporting driver alertness without introducing distraction. Participants’ alertness was assessed using the Karolinska Sleepiness Scale (KSS), alongside post-drive interviews to evaluate perceived usefulness, naturalness, and engagement.

This study contributes:
\begin{enumerate}
    \item One of the first field studies using a real CA, deployed in real-time, during \textit{authentic} automated driving, demonstrating its capacity to mitigate passive fatigue and support driver alertness under highly ecologically valid conditions.
    \item Thematic user insights into drivers’ experiences and perceptions, highlighting the factors that shape the acceptability of CAs in safety-critical contexts.
    \item Emerging user archetypes that align with established CA design frameworks, revealing how safety-first, entertainment-seeking, and socially oriented drivers value different aspects of conversational support.
\end{enumerate}

Together, these contributions advance our understanding of how conversational interfaces can be designed to balance safety and engagement in highly automated vehicles. More broadly, they point to the value of archetype-sensitive design in shaping future human–AI mobility systems.

\section{Related Work}

\subsection{Designing against Passive Fatigue in Conditional Automated Driving}

In conditional automated driving, drivers’ alertness may degrade during monotonous driving tasks. Prior work identifies two key dimensions of monotony, road design monotony and roadside variability, which can rapidly reduce driver alertness \cite{Larue2011-vp}. Such degradation widens the readiness gap between passive monitoring and active control \cite{Lee2019-gf}. This gap has motivated growing interest in Human–Machine Interface (HMI) strategies that can subtly reorient driver attention in preparation for potential transitions \cite{Baldwin2019-hu}.

A variety of HMI solutions have been proposed, including explicit visual, auditory, and haptic alerts \cite{van-Veen2017-we, Nees2016-fg, Khorsandi2023-kg}. However, these modalities can disrupt the user experience, particularly when delivered during pre-transition periods where readiness is low but a takeover is not yet required. In response, researchers have increasingly explored “gentle” interaction methods in authentic driving contexts \cite{yu2025-nd}, including gamification and CAs.

Research has examined how playful or game-like elements can help keep drivers alert by adding entertainment value. \citet{Steinberger2017} demonstrated how the inclusion of gamified elements can sustain driver attention during simulated driving. Introducing speed-control feedback, challenges and a score-based system was found to alleviate driver boredom and encourage engagement with the task. \citet{Bier2019-xr} also tested a gamified in-vehicle interaction system, asking trivia questions linked to the driving simulator task. They compared three groups: a control, a co-driver condition, and the gamified system. Results showed that both the co-driver and the gamified systems helped delay the onset of fatigue. The gamified system also led to better engagement over time. 

These results point to the potential of leveraging lightweight and joyful interaction to maintain attention and engagement. Game-like tasks can increase cognitive load just enough to prevent boredom without overwhelming the driver. However, they also present design challenges. Systems need to be simple enough not to distract drivers from takeover events, but engaging enough to be meaningful.

\citet{Mahajan2021-rl} used a WoZ design to test whether a simple voice agent could reduce passive fatigue during a 30-minute L3 drive. The agent provided hazard alerts, mock reminders, and general conversation using pre-recorded messages every three minutes. Compared to driving without the agent, participants showed reduced sleepiness and no microsleeps. A follow-up study reported that agent-assisted participants were 39\% more likely to complete a successful takeover \cite{Mahajan2021-rl}. This suggests that conversational interactions may both support alertness and improve takeover performance. \citet{large2016stimulating} reported similar effects in their WoZ study. 

More recently, researchers such as \citet{Huang2024-qe} have begun introducing real-time LLM-powered CAs into a simulated driving task. Their system prompted drivers to engage in frequent, simple conversations during a long, monotonous session. High- and low-frequency conversations elicited greater electroencephalography (EEG) arousal than the absence of interaction. This study was one of the first to apply a genuine large LLM in a simulated driving context. 

Most prior studies addressing passive fatigue, regardless of their intervention, used driving simulators, which do not fully reflect what happens when drivers are in motion \cite{large2016stimulating, Huang2024-qe} on a road and are responsible for their own safety. Simulators are helpful for early testing, but do not capture all the cues that shape how people feel about in-vehicle technology. Studies have also shown that the presence or absence of vehicle motion itself can significantly affect driver responses \cite{Motion-Sadeghian_Borojeni2018-gh}.  Therefore, while past work shows that CAs and gamification can help interrupt passive fatigue, the effectiveness of playful, LLM-powered CAs in addressing passive fatigue\textit{ in a real-world driving context }remains unknown. 

Our study addresses this gap with one of the first deployments of a fully-functional LLM-powered CA in a genuine L3 automated vehicle prototype during monotonous driving. The CA was designed to provide a natural, seamless and playful form of safety-forward interaction. This approach enabled us to gain rich insights into how participants experienced the agent strategy, how useful they perceived it to be, and what changes they would expect for future deployment.

\section{Designing Zoe: A Conversational Agent to Combat Passive Fatigue}

We designed a CA, Zoe, specifically to help drivers of conditionally automated vehicles combat passive fatigue. At the time of the study, Zoe was powered by the latest available model (OpenAI GPT-4\footnote{https://openai.com/index/gpt-4-research/} with a real-time API), ensuring responsiveness and naturalness in interaction. The authors acknowledge that other LLMs, as well as more recent versions of GPT, have since become available and may produce responses that differ from those generated during our study.

The design drew on prior simulator-based studies of conversational agents, which emphasised the importance of low-complexity dialogue, natural speech, and short communication in reducing fatigue \cite{large2016stimulating, Huang2024-qe}. From a safety perspective, supporting situational awareness has been widely discussed in the context of L3 driving \cite{Endsley1995-ll, Capallera2023-ux}, with particular emphasis on context-aware information exchange for developing adaptive interfaces and providing implicit interaction \cite{Schilit1994-dr, Bolchini2007-as}. 

Building on these findings, Zoe was developed around three core elements: 
\begin{enumerate}
    \item \textbf{Engagement strategies} encouraged drivers to notice and describe their surroundings (e.g., "What can you see around us?") or comment on expected roadside wildlife.
    \item \textbf{Interaction management} protocols ensured the agent never assumed responses, paused if participants appeared busy, and included recovery strategies such as, "I apologise, could you please repeat that?"
    \item \textbf{Safety protocols} required the agent to regularly remind drivers that road attention always takes precedence and to immediately halt interaction if the CA became a distraction.
\end{enumerate}

In addition, light cognitive-weight gamification features were included to broaden appeal and encourage engagement. \textit{Zoe} was prompted to provide context-relevant prompts tied to road vigilance. This included optional micro-interactions drawn selectively from the Entertainment archetype (e.g., brief challenges, environment-linked facts).  The intention was to support attention and provide gamification elements \cite{Bier2019-xr} without impacting on safety goals. A little-shot prompting approach was used to enhance the user experience by reducing the risk of hallucination, improving the accuracy of personalised responses, and supporting greater stability, consistency, and clarity \cite{Mortezapour2025-ys}.

\textbf{Design goals}

\begin{enumerate}
    \item Safety first: safety prompts are primary; enrichment is opt-in.
    \item  Low cognitive buy-in: micro-tasks only (seconds, not minutes), with clear exits.
    \item Context fit: content ties to the drive (e.g., local context, vigilance cues) rather than free-form chatter.
\end{enumerate}

Goal order reflects a deliberate emphasis on cognitive safety before contextual richness. We therefore prioritised \textit{low cognitive buy-in} to minimise the risk of inducing active fatigue (inverse of passive fatigue) \cite{May2009-vc}. This emphasis ensured that any contextual dialogue remained secondary to maintaining attentional safety.

Once triggered, the CA operated fully autonomously, generating content in real time rather than relying on pre-scripted or Wizard-of-Oz responses. We relied on the live API to preserve authenticity and demonstrate how a deployable agent could operate in the vehicle. The feature was integrated into a bespoke iOS application (see Figure\ref{fig:meterial}-a), which allowed experimenters to discreetly position the device out of the driver’s view and interact via voice alone. The app supported pre-defined triggers based on time and GPS location, and enabled the input of participant names to deliver more empathetic, personalised dialogue.

The design was refined through approximately 10-15 pilot tests with the broader research team to confirm audio quality, interaction timing, and safety prompts. This allowed us to adjust the balance between naturalness and conciseness, ensuring conversations remained engaging without overloading drivers.

Examples of Zoe’s interaction included:
\begin{itemize}
    \item \textit{Naturalness}: "Nice to meet you, Mindy. I'm Zoe, your driving companion. How are you finding the drive so far? Are you feeling alert and engaged?"
    \item \textit{Context-awareness}: "Let’s keep an eye out together. Have you noticed any wildlife near the track?", "What do you see around us? Anything interesting catching your eye?"
\end{itemize}

\section{Method}
A between-subjects test-track study was conducted to investigate the impact of a real-time CA on passive fatigue in the context of conditional automated driving. Perceived usefulness, acceptability, and usability were also assessed via semi-structured interviews. Forty participants each completed a fifty-minute drive in an L3 prototype vehicle on a closed test-track. The chosen duration exceeds that of related studies that employed approximately 25-minute driving periods \cite{large2016stimulating, Mahajan2021-rl}. Additionally, prior work reports that fatigue (impaired vigilance, increased sleepiness and mind wandering) occurs after as little as 15~minutes of automated driving \cite{Vogelpohl2019-ap}.

Participants were assigned to one of two conditions: CA (received a short, spoken interaction with a real-time agent near the end of the drive) or Control (completed the same drive with no agent interaction). Following the drive, semi-structured interviews were conducted with both groups, during which the control group also experienced the conversational agent (though not in context). 

\subsection{Participants}

Forty participants were recruited (m:f = 24:16, \emph{M} = 56.8 years, \emph{SD} = 19.2). Recruitment was conducted via email invitations, social media posts, flyers, and snowballing, targeting diverse communities including campus advertisements, driving clubs, women’s community groups, senior centres, and public noticeboards. Participants completed a pre-screening form including demographic details and health-related questions and were provided with study information sufficient to understand potential risks. 
The sample included younger drivers and older drivers (65+) to examine potential age-related differences in driver states and reactions to technology. All participants held valid driver licences and at least two years of driving experience. Individuals reporting a history of fatigue or sleep-related medical conditions were excluded, and participants were instructed to avoid caffeine and other stimulants or depressants on the day of the session. Participants were allocated to either the CA group (\emph{n} = 25, m:f = 14:10, \emph{M} = 53 years, \emph{SD} = 20.2) or the control group (NCA) (\emph{n} = 15, m:f = 10:5, \emph{M} = 55.2 years, \emph{SD} = 17.2). Given fixed recruitment resources, we adopted an approximately 2:1 allocation favouring CA to maximise data from the condition of interest. 

All participants received information about the study, specifically a statement about protecting safety and privacy and signed informed consent and image release forms approved by the Ethics Committee of Queensland University of Technology (approval number 8522). Each participant received AU\$150 as compensation for completing the study. This was deemed appropriate due to the participants' on-track time commitment (about 2-3h, including waiting time), and significant travel time to and from the test track (approximately 1.5h).

\subsection{Apparatus}

\subsubsection{Vehicle and SAE Level 3 Automation}
\begin{figure}
  \includegraphics[width=\textwidth]{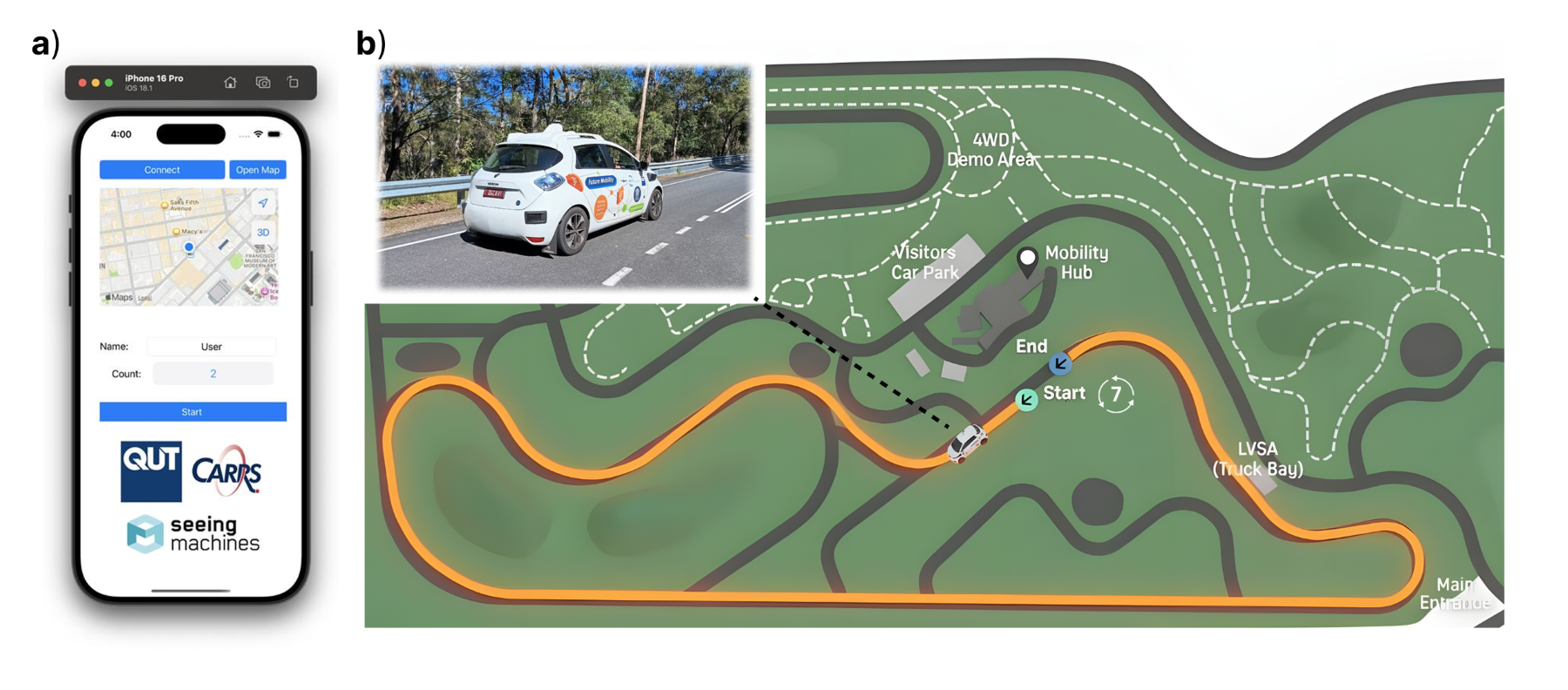}
  \caption{a) The iOS App developed for implementing the CA - Zoe; b) The L3 automated vehicle prototype, Zoe and closed-road circuit employed for monotonous automated driving trials.}
  \Description{Panel-a shows the experimental automated vehicle prototype used in the study to support L3 driving. Panel-b illustrates the layout of the closed test track where the automated driving trials took place. The orange path represents the fixed route along which participants completed seven uninterrupted laps. The circuit includes gentle curves and long straight segments, offering a repetitive and low-demand driving environment suitable for inducing passive fatigue under automated conditions.}
  \label{fig:meterial}
\end{figure}

The study was conducted using a Renault Zoe electric prototype vehicle adapted for SAE Level 3 (L3) automated driving (Figure\ref{fig:meterial}-b). Conducting the experiment in L3 facilitated exposure to authentic vehicle motion, vibration, road curvature, and naturalistic fatigue onset patterns. This methodological choice meaningfully enhances ecological validity and distinguishes this study from prior CA-focused fatigue mitigation research \cite{large2016stimulating, Large2019-ou, Mahajan2021-rl}. The vehicle operated in automated mode on a closed-road circuit. A trained safety driver was seated alongside the participant and could assume control at any time, ensuring safe execution of handover and takeover procedures.

\subsubsection{Implementation of the Conversational Agent}

The CA was deployed as a mobile phone application, programmed to initiate a 120-second dialogue \cite{Saxby2008-mu, Mahajan2021-rl} at the start of the sixth monotonous lap on the track. Audio was delivered through bone-conduction headphones. This enabled participants to hear both the agent and ambient vehicle sounds without interference. Participant responses were captured via the headphone’s built-in microphone, creating a natural, hands-free conversational experience. All conversations were transcribed and stored locally on the device. The prompt development and interaction strategy adopted a LLM-based agent design for combat fatigue based on \citet{Yu2025-nl}'s design. For the control (NCA) group, the agent was demonstrated only during the post-drive interview to avoid influencing their in-vehicle experience or introducing carryover effects between conditions. The purpose of this demonstration was to give NCA participants the opportunity to experience the CA after the experiment, allowing them to imagine how this form of interaction might have applied to their own automated drive they had just experienced, and to reflect on this during the post-drive interview
 (see Appendix~\autoref{table:p_1}).

\subsubsection{Side Video Recording}
In-car video was recorded using an iPhone mounted on the safety driver’s side window. The camera captured the driver’s behaviour, such as posture, attention, and interactions throughout the drive (\autoref{fig:teaser}). Recording began after the familiarisation phase and concluded once the participant completed all driving trials. This provided a continuous behavioural record to support analysis of engagement and fatigue.

\subsubsection{Karolinska Sleepiness Scale (KSS)}
The KSS was administered to capture subjective states of passive fatigue during the monotonous driving trial. Widely used in fatigue research \cite{large2016stimulating, Mahajan2021-rl}, the KSS provides a 1–9 scale for participants to self-report their level of vigilance. After the familiarisation phase, the safety driver explained the scale to participants and asked them to report their ratings verbally at predetermined points during the drive (see Section \ref{sec:monotonousDriving}). These were recorded by the safety driver. These scores served as a complementary indicator alongside behavioural observations and interview data. KSS administration may introduce brief attentional resets, potentially mitigating deeper fatigue onset. To minimise this effect, researchers rehearsed KSS ratings twice with participants before the driving trial, and during the drive, KSS ratings were elicited with brief, consistent prompts across conditions.

\subsubsection{Semi-structured Interview}
Post-drive, participants completed a semi-structured interview with open-ended questions (see Section \ref{sec:post-drive}). Interviews were audio-recorded using a digital pen, with one researcher leading the discussion and another taking field notes. For the NCA group, the agent was demonstrated at this stage by the assisting researcher to elicit comparative feedback.

\subsection{Procedure}

\begin{figure}
  \includegraphics[width=\textwidth]{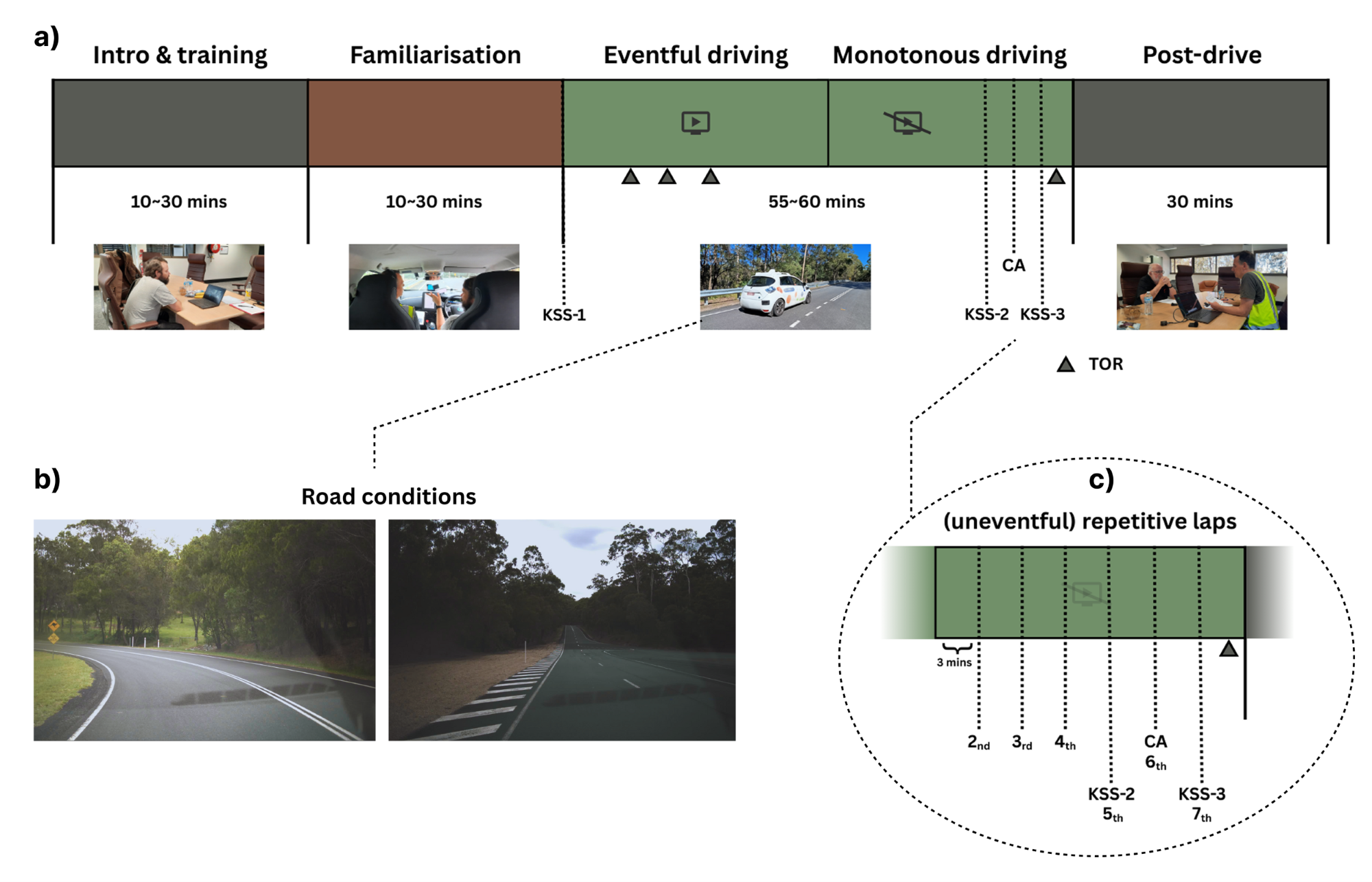}
  \caption{a) Timeline of the experimental protocol, b) the road conditions and c) the details of the monotonous, repetitive laps.}
  \label{fig:procedure}
  \Description{Sub-figure a shows a timeline diagram of the experimental protocol. Four stages are displayed from left to right: introduction and training, familiarisation, eventful driving，monotonous driving, and post-drive. Each stage includes the approximate duration and an accompanying photo. Indicators of when Karolinska Sleepiness Scale ratings were taken, when the conversational agent was activated and when the NDRT was turned off are also presented. Sub-figure b presents two road condition example screenshots at the bottom showing the environment of the test-track. Sub-figure c explained the details of the uneventful repetitive laps.}
\end{figure}

The study was conducted during daylight hours (08:40–17:00). Each participant completed an on-road test-track session in the AV. The session was structured into four phases: Introduction and Training, Familiarisation, Monotonous Driving, and a Post-Drive Interview (Figure\ref{fig:procedure}-a). During the drive, the CA either initiated an interaction with the driver at a predetermined location or remained inactive, resulting in continued monotonous driving.

\subsubsection{Part I – Introduction and Training}
Participants were welcomed with a briefing and signed informed consent forms. An introductory video presented the study aims broadly as exploring interactions with a state-of-the-art digital CA during automated driving. The video also described system functions and safety protocols, emphasising that an intelligent conversational agent may initiate a short dialogue during the drive. Participants were instructed not to converse with the safety driver once the trial commenced to avoid confounding fatigue measures and the effect of conversation with the CA. This included instructions to not check their phone, watches or interact with the central console. They were also briefed on their responsibilities as conditionally automated drivers, specifically, that they were permitted to remove their hands from the steering wheel and engage in limited non-driving-related activities, but were required to remain ready to take over control when prompted. This training included a standardised instructional video that had been used in previous studies for the experimental AV.

\subsubsection{Part II – Familiarisation}
Participants were guided from the preparation area to the start of the track. To build comfort, trust in the AV and prevent mode misidentification, the safety driver explained the vehicle’s functionality and repeated key safety instructions. Participants adjusted the seat and performed a 10-minute warm-up drive, including exposure to automated driving, manual control, and transitions of control (from manual to automated and vice versa).

\subsubsection{Part III – Monotonous Driving}
\label{sec:monotonousDriving}
Participants then undertook a prolonged driving trial, lasting approximately 55 minutes in total. This driving trial included: 1) Thirty minutes of more eventful automated driving with three planned takeover requests (TORs) while watching videos as an NDRT (all participants watched the same series of short mini-documentaries regardless of condition) and taking different laps and turns around the track. TORs were provided to contextualise the interaction with the CA with respect to mitigating passive fatigue in the context of L3 automated driving, where drivers may be requested to take over. This contextualised the CA as a potential safety feature in the post-drive interview.
2) Followed and contrasted by multiple repeating laps, specifically designed to be passively fatiguing by being uneventful, low-stimulation and repetitive; and
3) A takeover request in the last lap, further emphasising the conditional automated driving context for the post-drive interview.
Each lap included straight segments, wide curves, and sharp turns. The vehicle operated in automated mode at a maximum speed of 50 km/h, as required by local regulations, on a single carriageway with two lanes (Figure\ref{fig:procedure}-b).

In this session, participants first engaged in a non-driving-related task (watching video clips). The video was then turned off by the safety driver, leaving participants idle and without purposeful activity. This was followed by seven repetitive laps of a closed, low-stimulation circuit (each ~3 minutes), designed to induce passive fatigue through monotony (Figure\ref{fig:procedure}-c). During the penultimate lap, the CA initiated a 120-second dialogue \cite{Mahajan2021-rl} at a fixed location on the track, prompting drivers to observe and verbally comment on surrounding road and environmental features. Participants in the control condition completed the same trial but without CA activated. A final takeover request due to an emergent lane-changing event was issued during the last lap to re-contextualise the CA in the context of conditional automated driving.

The KSS was administered for driving at the following points:
\begin{itemize}
    \item Pre-drive (baseline),
    \item Prior to the activation of the CA (on the start of the 5th monotonous lap, across both conditions),
    \item Following the CA interaction (on the start of the 7th monotonous lap, across both conditions).
\end{itemize}


\subsubsection{Part IV – Post-Drive}
\label{sec:post-drive}
After the drive, participants took part in semi-structured interviews (15–30 minutes) in a quiet preparation room. The interviews explored their subjective experiences of automated driving with or without the CA and probed perceptions of alertness, naturalness, and engagement. Participants in the control group were also given the opportunity to experience the CA, although not within the driving context.

The interview design followed a structured progression:
\begin{itemize}
    \item \textit{Warm-up} – Participants answered introductory questions about their background and general driving experience.
    \item \textit{Initial reflections} - Participants were first invited to share their overall reflections on the CA (Q1). Participants is in the control group completed a demonstration with the CA at this point. 
    \item \textit{Disclosure and perceived effects on alertness} - The researcher clarified the design intent of the CA. Participants discussed whether and how the agent influenced their state of alertness, providing specific examples of situations where it was helpful or unhelpful (Q2\&3).
    \item \textit{Interaction experience and enjoyment} - Participants described the naturalness and challenges of conversing with the agent, as well as its entertainment value and impact on enjoyment of the drive (Q4\&5).
    \item \textit{Willingness to use and improvements} - Finally, participants reflected on their likelihood of using such a system in the future and suggested potential design enhancements (Q6\&7).
\end{itemize}

All interviews were audio-recorded with consent and transcribed verbatim. Observational fieldnotes were collected during and after each session to capture behavioural responses and contextual details.

\subsection{Analysis}
The analysis process comprised three interconnected stages: (1) video analysis of in-car recordings, (2) quantitative assessment of KSS ratings, and (3) thematic analysis of post-drive interviews. These stages were not conducted in isolation; rather, each iteratively informed the others to strengthen interpretation. This approach adapted a concurrent nested analysis, privileging qualitative analysis \cite{creswell2010sage} while using quantitative measures in a complementary role to support interpretation.

\subsubsection{Video Analysis of In-Car Recordings}
To complement the post-interviews, in-car video recordings were examined to capture observable indicators of attention and driver states that may not have been explicitly reported by participants. The analytical focus was on identifying the types of behaviours present within the observation windows (i.e., during CA interaction or the NCA periods), rather than quantifying their frequency, acknowledging that some behaviours may vary across individuals. Audio-visual data analysis is increasingly employed in HCI and CSCW research to surface ‘seen-but-unnoticed’ aspects of interaction \cite{Heath2010-xw, Brown2023-bg}. Following this, we reviewed 40 in-car side video recordings (each approximately 55 minutes). Three participants were excluded as the video was cut due to the recording phone overheating. Consequently, the analysis focused on 37 segments (approximately 8 minutes each) where the CA was active and compares them to corresponding periods without agent interaction.

For each instance, the analysis centred on fixed track locations where the agent was triggered. Observations were coded within a three-minute window before the conversational onset, a three-minute window during the intervention lap, and a one-and-a-half-minute window after the intervention lap (before the planned takeover, which is out of the scope of this paper). This window was extended when notable behavioural changes persisted. Any behaviours prompted during safety driver-initiated KSS probes were excluded from coding.

Particular attention was paid to body movements (e.g., gestures of thinking, communicating, or showing fatigue), facial expressions (coded using an emotional wheel), and head or eye behaviour (e.g., gaze direction, noticeable changes in blink rate). These cues provide insights into situation awareness \cite{Endsley1995-ll}, arousal \cite{McWilliams2021-lz}, and potential states of passive fatigue \cite{Brown2023-bg}. 

To support collaboration, coding was documented in a shared Google Sheet, enabling authors to refine categories and resolve discrepancies. However, all video playback was conducted locally on a secure data drive to maintain participant privacy and data protection. The resulting patterns or insights were then cross-referenced with interview data to explore not only what participants said about their experiences, but also how their behaviour unfolded in situ.

\subsubsection{Karolinska Sleepiness Scale}
Concurrently, participants’ subjective sleepiness ratings on the KSS were analysed to provide quantitative support for the findings. Descriptive and statistical analyses were conducted to examine user differences in KSS scores. Two aspects were of particular interest: (1) whether individuals’ KSS scores changed before and after interacting with (or without) the CA, suggesting the agent’s potential to mitigate fatigue; and (2) whether these changes aligned with behavioural patterns identified in interviews and video analysis, indicating consistency between subjective reports and observed behaviour. In this way, KSS results offered an additional lens to triangulate and substantiate the themes.

\subsubsection{Thematic Analysis of Post-Interview}
The interviews were analysed using notebook Qualitative Content Analysis (QCA) procedures \cite{Kuckartz2019-ae}. Transcripts were generated with the support of Feishu Minutes and the Whisper API, then manually verified word-for-word by two authors. In total, 39 transcripts were analysed; one participant was excluded for not consenting to the use of AI-enabled transcription tools. The two authors first familiarised themselves with the transcripts before applying a deductive coding approach. Prior to coding, relevant theoretical perspectives for passive fatigue were discussed. This includes the the Yerkes–Dodson law of the relationship between cognitive load and performance \cite{McWilliams2021-lz} and various technology acceptance models \cite{Venkatesh2008-ep, Hutchins2017-ov, Chen2023-hz}. 

All transcripts were coded in MAXQDA \footnote{https://www.maxqda.com/}. Codes captured participants’ reflections on their experience, attitudes, interactions and recommendations with both the automated driving system and the CA. To ensure rigour, Authors 1 and 2 independently coded an initial 20\% of the data. Inter-coder reliability was calculated and indicated strong agreement (Cohen\'s k = 0.84). Discrepancies were resolved through discussion and iteration before proceeding.

In subsequent rounds, the remaining interviews were divided between the two authors, with regular meetings to iteratively refine and consolidate the codebook. Emerging codes were organised into initial themes, which were further discussed with the wider research team. Insights from video analysis and KSS results were later used to challenge, support, and deepen the thematic interpretations.

Two authors jointly conducted the video coding, KSS analysis, and thematic interpretation. These analyses were carried out in parallel, with thematic analysis iteratively informed by insights from the video and KSS results. Overall, the video data revealed how drivers behaved when interacting with the CA, the KSS captured their subjective states before and after the intervention, and the thematic analysis investigated perceptions of usage and effectiveness through participants’ accounts. Together, these sources of evidence were triangulated to refine and interpret the findings.

\section{Results}

\subsection{Drivers’ Behaviour while Driving}

\begin{figure}
  \includegraphics[width=0.9\textwidth]{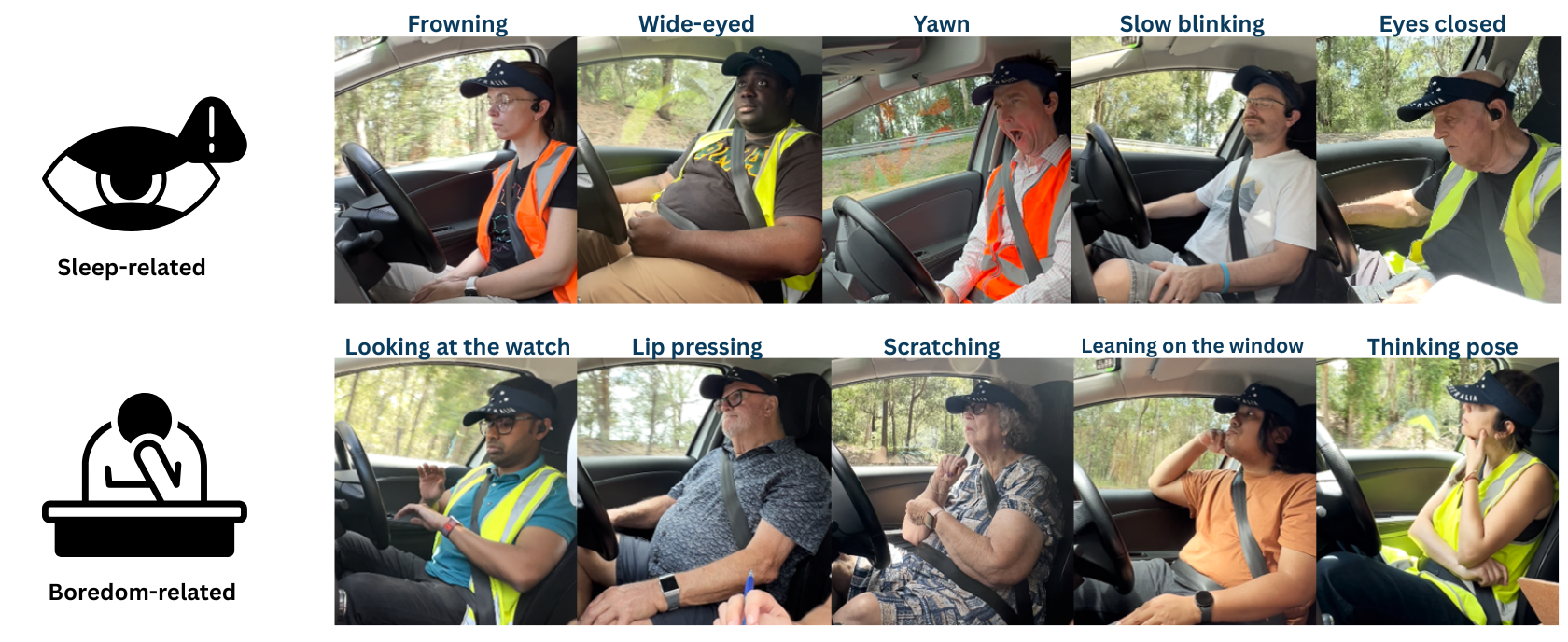}
  \caption{In-situ fatigue-related behaviour examples from participants in L3 driving context.}
  \Description{The photo montage shows participants in the automated vehicle displaying observable behaviours linked to passive fatigue. Sleep-related indicators include frowning, wide-open eyes, yawning, slow blinking, and eyes closed. Boredom-related indicators include looking away, lip pressing, scratching, leaning on the window, and a thinking pose.}
  \label{fig:video_results}
\end{figure}

\autoref{fig:video_results} summarises the common behaviours observed, along with any unexpected driving behaviours.

\textbf{Fatigue-related behaviours} 
During the observational window prior to the intervention, participants frequently displayed behaviours linked to drowsiness. These included yawning, slow blinking, and partial eye-lid closure. Instances of wide-open eyes or frowning also indirectly indicated attempts to fight sleepiness. Following the CA intervention (CA group), some instances of yawning and slow blinking persisted.

Small, restless movements were common and interpreted as signs of boredom. Examples included looking at nearby objects (such as the side door, a watch or the centre console), and oral activities like licking lips, lip pressing, pouting, self-talk, or whistling. Hand movements were frequent, such as scratching the neck, thighs, or arms; tapping rhythmically on a palm or thigh; twirling fingers; and touching the eyes, face, hair, forehead, or nose. Postural shifts were also observed, including crossed arms, a thinking pose, and leaning on the window. Some combined behaviours, such as adopting a thinking pose while looking upward or fast blinking, suggested that attention was directed away from driving-relevant information. During and after the CA interaction (within the coding windows), these small movements decreased.

Overall, these observations illustrate how drivers behaved in situ during monotonous L3 driving, revealing both drowsiness-related signs and boredom-induced activity.

\begin{figure}
  \includegraphics[width=0.9\textwidth]{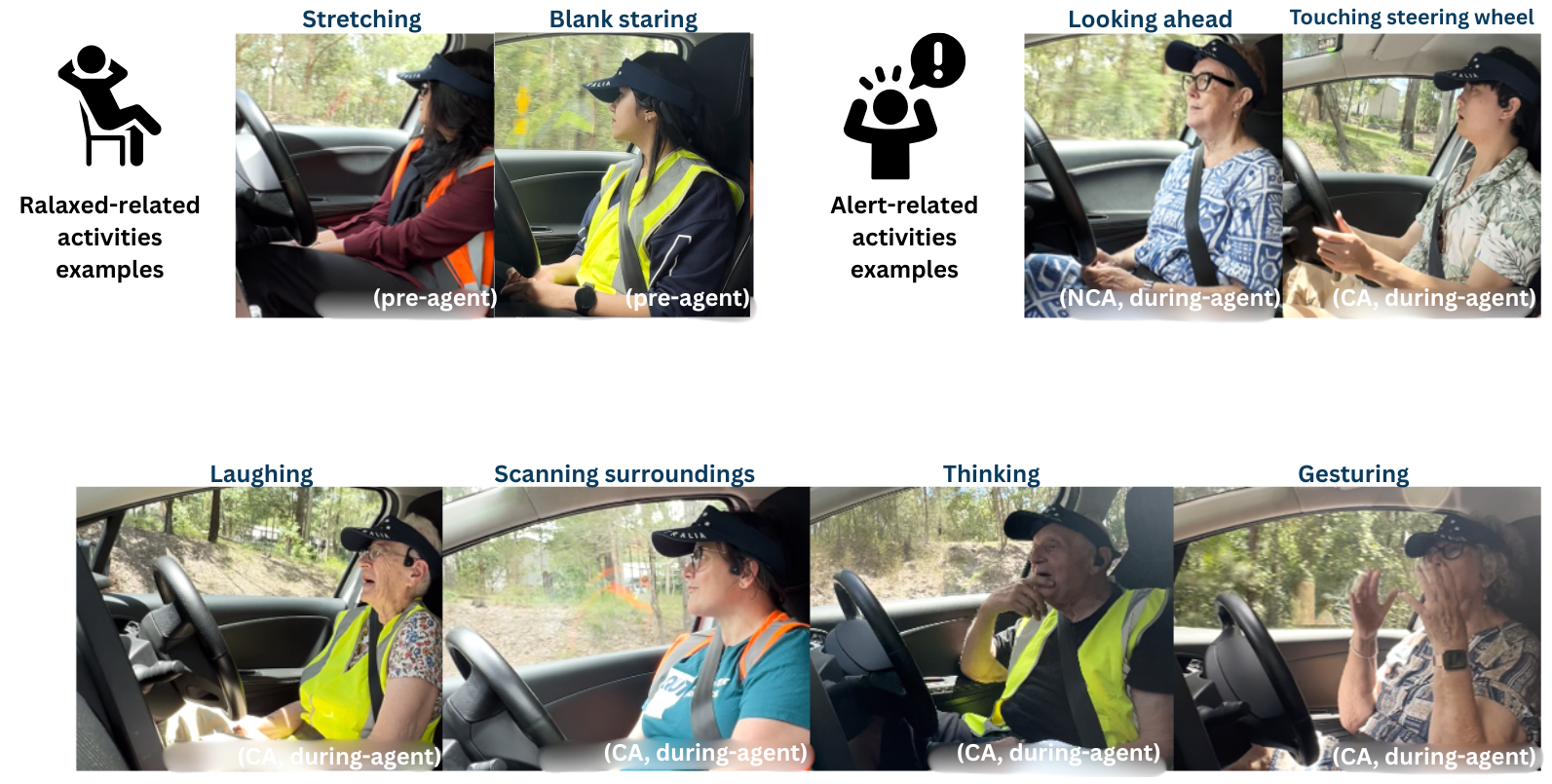}
  \caption{Examples of passive, relaxed states and active, engaged behaviours during the monotonous drive and when interacting with the conversational agent.}
  \Description{The photo montage shows participants in the automated vehicle displaying observable behaviours linked to passive fatigue. Sleep-related indicators include frowning, wide-open eyes, yawning, slow blinking, and eyes closed. Boredom-related indicators include looking away, lip pressing, scratching, leaning on the window, and a thinking pose.}
  \label{fig:video_results2}
\end{figure}

\textbf{Relaxation and alert.} 
Both relaxed and alert behaviours were observed, appearing before and after the CA intervention (\autoref{fig:video_results2}). Relaxed behaviours included stretching the arms, shoulders, or legs; slightly tilting the head upward; leaning against the window; looking around with broad head movements; and adjusting posture in the seat.

Some participants appeared to remain alert, as shown by extended periods of looking straight ahead or scanning the surroundings while maintaining an upright posture. One participant stated during the post-agent KSS probe, “I won’t trust the system,” to the safety driver. However, relaxation and alertness were not mutually exclusive. Drivers who kept their gaze forward for relatively long periods also occasionally displayed relaxed or fatigued behaviours. In the NCA group, one participant temporarily took control to avoid roadside wildlife but soon reverted to behaviours reflecting tiredness and relaxation.

These observations suggest that relaxation and alertness were dynamic states, potentially shaped by trust in the AV or by driving experience or habit (see thematic section).

\textbf{Engagement and attraction.} 
Compared with the NCA group, CA participants displayed positive affect during and after conversations (\autoref{fig:video_results2}). Facial expressions such as smiling, laughing, or showing surprise were observed, and some participants joked with or were amused by the agent. Following the framework of \citet{Plutchik1982-cw}, these expressions were coded as Joy and Ecstasy. Others engaged less, responding only briefly. For example, one participant rejected the CAs opening gambit, replying; “Not really… but thanks for asking, you keep me awake.” Down-turned mouth corners were also identified as micro-expressions, potentially signalling dissatisfaction with the content.

Participants increased their scanning of the surroundings and gave verbal responses in response to the agent's prompting. Cognitive activities during conversation were also observed, such as lateral eye movements and thinking pose. Some also used hand gestures while speaking, typically between the steering wheel and the head. One participant instinctively touched the steering wheel when the voice was first activated, then relaxed and engaged once realising it was not a takeover request.

Participants also expressed varied communication needs. Some requested functional features, such as \textit{“Can you play some music?”}, while others sought entertainment, asking \textit{“Let’s play a word game.”} After the dialogue ended, a few attempted to re-engage Zoe with prompts like \textit{“AI agent, what’s your name?”}. This indicates a desire to continue, though they were unaware that the system was programmed to stop after a fixed location.

The above behaviours were not observed in the NCA group. Overall, conversations with the CA elicited positive emotions associated with joyfulness, combined with gaze behaviours and hand movements. Participants expressed willingness to engage with the CA, though their needs and expectations varied widely.

\subsection{Self-reported Sleepiness Scale}

A Mauchly’s test was firstly conducted, indicating that the assumption of sphericity was met, $\chi^{2}(2)=0.84, p=.656$.  Thus, a 2 (Group: CA vs. NCA) $\times$ 3 (Time: Baseline, Pre, Post) mixed-design ANOVA was used to examine participants’ KSS ratings. The analysis revealed a significant main effect of \textit{Time}, $F(2,76)=9.93, p<.001, \eta^{2}=.21$, and a significant \textit{Time $\times$ Group} interaction, $F(2,76)=5.51, p=.006, \eta^{2}=.13$.  The main effect of \textit{Group} was not significant, $F(1,38)=0.32, p=.575, \eta^{2}=.01$. 

\begin{figure}
  \includegraphics[width=\textwidth]{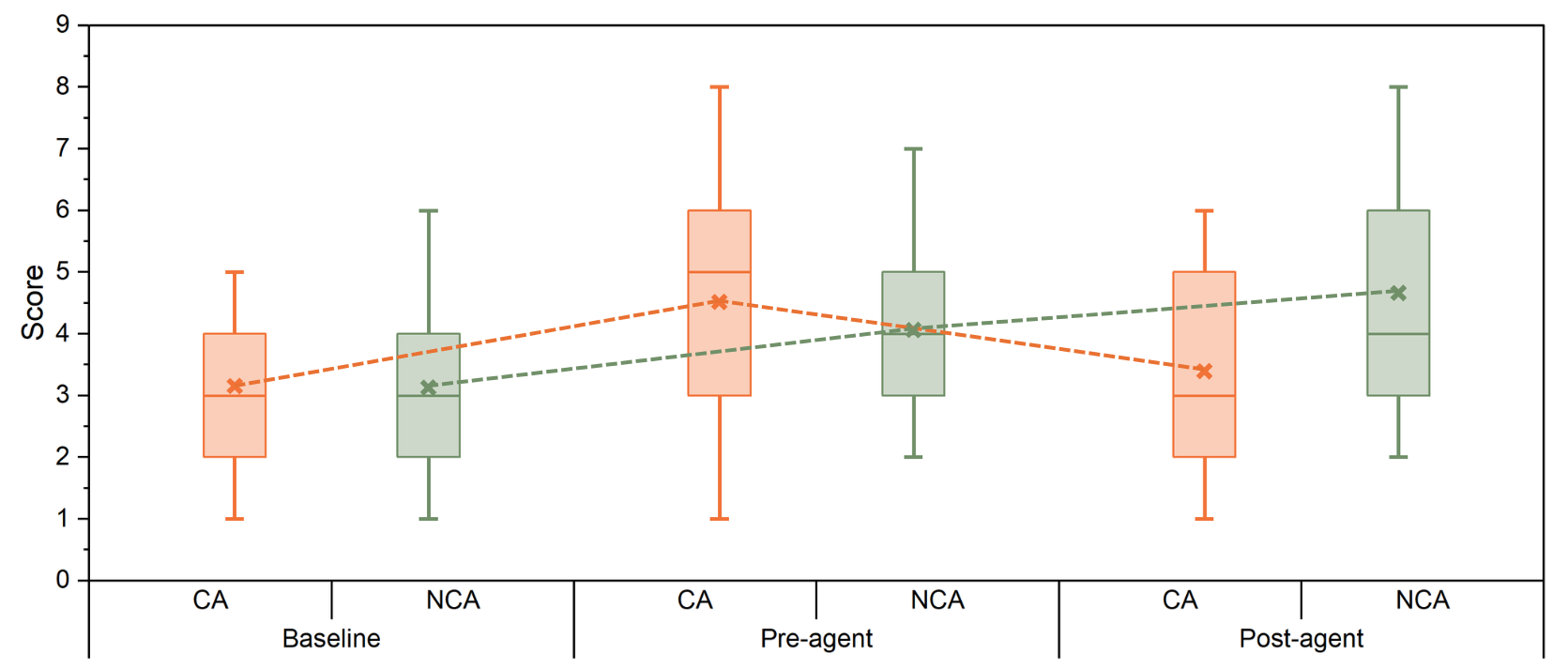}
  \caption{Karolinska Sleepiness Scale (KSS) scores at baseline, pre-interaction, and post-interaction.}
  \Description{The figure presents boxplots of Karolinska Sleepiness Scale (KSS) scores at three time points: baseline, pre-agent, and post-agent. Each time point is shown separately for participants who interacted with a CA (CA, orange) and those who did not (NCA, green). KSS scores are presented on the vertical axis, time-points are presented on the horizontal axis.}
  \label{fig:kss}
\end{figure}

The results indicate that participants’ KSS scores overall differed across the three \textit{Time} points. In general, the KSS score increased from Baseline ($M=3.15, SD=0.24$, $p<.001$) to Pre ($M=4.29, SD=0.32$) and decreased from Pre to Post ($M=4.03, SD=0.28$, $p=.003$). When considering the interaction between Time and Group, the results revealed divergent trajectories (\autoref{fig:kss}). For the group with CA, the KSS measured in Pre was significantly higher than in Baseline ($p<.001$) and Post ($p=.002$). For the NCA group, the KSS in Post was significantly higher than the Baseline ($p=.004$). The difference between CA and NCA in KSS was only significant in Post ($p=.028$).

These results indicate that passive fatigue increased over time in the absence of an intervention (NCA). However, engaging with the CA appeared to reverse this trend. At the two earlier time points, before the onset of L3 automated driving (Baseline) and before the interaction phase (Pre-agent), the CA and NCA groups did not differ significantly in drivers' levels of alertness. In contrast, following the intervention phase (Post-agent), the alertness levels of drivers in the CA and NCA groups diverged. As hypothesised, participants reported increased alertness following the CA interaction.

\subsection{Thematic Analysis}

Thematic analysis was used to 1) answer research questions regarding user perceptions of the CA as a useful safety tool, and 2) factors that affect the effective design of a safety-oriented CA for conditionally automated driving. The following sections will describe findings relating to the user experiences of both of these aspects.

\subsubsection*{\textbf{Theme 1. Perceptions of Usefulness for Managing Passive Fatigue}} Extending on observations and measures of vigilance, we explored how drivers perceived the CA as influencing their alertness. Three distinct themes within the overall perceived usefulness of the CA emerged: (1) interrupting sleep-related processes, (2) interrupting boredom-related processes, and (3) supporting alertness and engagement. These align with the core symptoms of passive fatigue presented by \cite{May2009-vc}.

\textbf{Interrupting sleep-related processes} Participants described the CA as useful and effective for interrupting sleep-related processes during automated driving. This was consistent across age, gender, condition and perceptions of usability of the agent. This finding corroborated with KSS scores, suggesting the CA was useful for waking participants up. Quotes of the following sentiment were nearly unanimous across participants. 

\begin{quote}
    \textit{"Definitely felt a level of awakeness, kick in, which is good."} -P12
\end{quote}
\begin{quote}
    \textit{"I felt sleepy, then this voice may alert me again when driving."} -P02
\end{quote}
These quotes effectively demonstrate the potential of the agent to interrupt sleep-related processes. Thus some drivers found that the agent useful for interrupting sleepiness but still found the interaction boring or distracting.

\textbf{Interrupting boredom-related processes} Some drivers saw the agent as breaking monotony or adding entertainment, others found it dull or superficial. Drivers who found the agent entertaining were often excited about the potential of such an agent in the future.

\begin{quote}
    \textit{"I think it would be positive, um, because... it's asking you questions and you're responding, it's sort of getting you to refocus and start using your brain differently again.}" -P95
\end{quote}

These users were also likely to see great potential in future iterations of similar agents: 
\begin{quote}
    \textit{"So it's certainly entertaining. And I think potential is boundless, it's just really cool".} -P01
\end{quote}


In contrast, some drivers found the interaction to be boring and unlikely to keep them engaged with the agent for an extended period of time. This was unrelated to their perceptions of the agent as a safety unit.

\begin{quote}
    \textit{"It was boring. It was... alerting, but, yeah, really boring."} -P04
\end{quote}
\begin{quote}
    \textit{"That's not what I call entertainment."} -P08
\end{quote}

While many participants offered suggestions for improving the entertainment value of the system, some stated that they would neither want nor need entertainment from such an agent. This was often accompanied with an explanation that human–agent conversation in any form simply did not align with their preferred forms of in-car entertainment. 
\begin{quote}
 \textit{"I'm sure they [other users] would [be entertained by the CA]. I'm sure some people would have lots of fun with it, really."} -P80
\end{quote}
\begin{quote}
    "\textit{That would fit some people. I don't know I really enjoy that." } -P49
\end{quote}
However, these users who did not enjoy the gamified aspects of the interaction did accept the agent as a necessary safety tool for AV driving. 
\begin{quote}
    \textit{"I wouldn't want to talk politics or anything, but yeah, just as an alertness thing now every, so if you do a long hour long drive, maybe every 20 minutes."} -P99
\end{quote}

\textbf{Supporting Alertness and Engagement} The data also revealed strong support for the CA as an alertness tool that actively re-engaged drivers with their surroundings.
\begin{quote}
    \textit{"It made me refocus. ... it reminded by asking about the signs, it sort of gets you to focus back on the, on the narrow scope of the road."} -P03
\end{quote}
This was also framed as providing a passenger-like presence that brought drivers back 'in-the-loop'. 

\begin{quote}
    \textit{"It’s kind of like having that… you bring a passenger along… less jarring than an auditory tone alert."} -P94
\end{quote}
\begin{quote}
    \textit{"It actually made me pretty alert… switched from a bit dizzy mode to a proactive mode because I have started conversation with a person."} -P08
\end{quote}
However, some drivers conveyed that the agent limited their ability to focus on the road. This idea was more common in drivers who did not respond to the fatigue intervention (KSS scores did not increase from Baseline to Pre-agent). Distraction was one of the most common suggestions among \textit{non-responders}:

\begin{quote}
    \textit{"I feel like it }[the CA]\textit{ could be a good thing or also a bad thing at the same time, because when you're driving, you try to make sure that you don't have a lot of distraction."} -P31 (NCA)
\end{quote}
\begin{quote}
    \textit{"A distraction ... until you got used to it. And then asking, what do you see around you? Some people might take their eyes off the road, literally, to have a look around. I think that might be a bit of a distraction.} -P79 (NCA)
\end{quote}

Even when participants were immersed in prolonged conditional automated driving, their responses were often framed in relation to manual driving rather than hypothetical future automated driving. This suggests that perceptions of safety-oriented CAs may be affected by mode awareness. Such awareness can influence the extent of driver engagement and, in some cases, lead the CA to be perceived as a potential distraction.
\begin{quote}
    \textit{"I already know what I'd do to help me stay engaged on a long drive, but then again, it wasn't automated, so I honestly don't know. So I normally just put on a podcast or music or something like that, and that's enough to keep me engaged." -P12}
\end{quote}
\begin{quote}
    \textit{"Yeah, it can be great ... but, if I'm in the manual mode and I have to concentrate on a lot of stuff on the road, sometimes it can be distracting. "} -P25 
\end{quote}

In conclusion, participants perceived the CA as useful in an AV context for reducing fatigue. However, these effects appeared to be shaped by multiple factors, including the content of the CA’s dialogue, individual characteristics such as personality and driving experience, the degree of agent type alignment with user needs, and contextual influences such as mode awareness.

\subsection*{Theme 2. Design and Interaction Factors}
This theme summarises the factors identified by participants as influencing the adoption of a CA. These factors centred on three key areas: user acceptance, accessibility, and usability. Each requires careful consideration by designers to ensure the system is appropriately adapted for conditional automated driving safety scenarios.

\textbf{Perceived Benefits Underpinning Acceptance}
Despite unanimous agreement that the agent was helpful as a safety tool and increased alertness, acceptance varied considerably in terms of whether participants would want such a safety feature in their own vehicles. Some were enthusiastic despite the technological limitations associated with the experimental nature of the agent's capability at the time of the study. Reported benefits included entertainment (e.g., playing games with the agent), safety (e.g., drawing attention or seeking a coffee replacement), and companionship (e.g., reducing loneliness).
\begin{quote}
    \textit{"I didn't like the game that we played, right? But yeah, the concept of it popping in and being available to talk through different things or play different games for that kind of thing is really good."} -P16
\end{quote}
\begin{quote}
    \textit{"It's going to sort of make me look about a bit more, take notice more of what's occurring... you know, the direction we're going in or signs or whatever... I didn't have any issues with that."} -P89
\end{quote}
\begin{quote}
    \textit{"It would be quite beneficial, I think, to have something like that if you were just travelling alone,"} - P80 (NCA)
\end{quote}

By contrast, others emphatically rejected the system for a variety of reasons. Prominent concerns were that some drivers simply disliked conversing with a machine.
\begin{quote}
    \textit{"I just find it weird engaging with an LLM… you don’t really get anything out of it."} -P12
\end{quote}
\begin{quote}
    \textit{"With the conversational agent, yeah, I don't know. I just found her like a computer box... which I found annoying"}. -P84
\end{quote}

A distinct group of participants also emerged who suggested they accepted the agent but only with certain provisions and only as a safety tool. 
\begin{quote}
    \textit{"Just as an alertness thing… every 20 minutes, two or three minutes, nothing more."} -P99
\end{quote}
\begin{quote}
    \textit{"Although I think that people will get annoyed very quickly with these things, so that would have to be, you know, in the regulations saying, like, they've got to check you every 10 minutes and you've got to respond in your voice on some sort of level or scales."} -P04
\end{quote}

Even among users who found the agent helpful for supporting their engagement with the driving task, acceptance pended on increased personalisation in future CAs. This idea applied to design cosmetics and presentation of the agent.
\begin{quote}
    \textit{"It's something I'd consider, but it would depend on, what the options with the car. Things like voice, I like to be able to customise to a different voice."} -P16
\end{quote}
Another strong theme was whether the agent was able to accommodate the users' unique desires, whims and nuances:

\begin{quote}
    \textit{"I think I would be more alert if it would be responsive to my desires to talk about some sort of a topic, not just being alert because you need to pick up some more wildlife, so what do you see? ... However, I found sort of a limit where it wants to stay in this alertness domain, right? So it doesn't want to talk about what's going on in US politics, right? It wants to stay there."} -P01
\end{quote}
\begin{quote}
    \textit{"You need a better AI than that... insufficient response to, I mean, it has a series of program responses, but the program is very small. You're not interacting with me in the true sense... I'm assuming that it would be programmable, this thing... you know, you have one that tells you a story, one that tells you a book."} - P113
\end{quote}
These quotes illustrate one of the strongest sentiments presented across users; that consistent future use would pend on the ability of the CA to shape to their preferences and accommodate their preferred engagement. This ranged from \textit{minimalist}, safety-oriented users who would prefer to only be disturbed for safety-critical reasons, users who preference all-in-one entertainment and integration to social users who desire meaningful and personal connection.  

\textbf{Accessibility-Related Barriers}

Although accessibility was not an initial focus, interview themes highlighted the need for CAs to be adaptable. Unlike prior simulator-based studies \cite{large2016stimulating, Huang2024-qe}, participants in the real-world setting not only discussed communication frequency and complexity, but also raised higher-level requirements affecting engagement\textemdash such as dialect, accent, and hearing-friendly dialogue.

\begin{quote}
    \textit{"Didn’t always pick up what I said… maybe my pronunciation." -P01}
\end{quote}
\begin{quote}
    \textit{"It's quite good because she's clear... She has a clear, sharp voice, good diction, and she doesn't speak too quickly... And in the world around me, people speak so quickly. My ears are 80 years old."} -P113
\end{quote}

\textbf{Usability-Related Challenges}

Participants also raised a range of usability concerns that may have shaped their willingness to adopt a CA. As expected, technical limitations, latency, and clunky error recovery were barriers to smooth use.

\begin{quote}
    \textit{"I will profess, I don't use Siri or anything else like that. So I also don't use ChatGPT either, so the whole concept of that sort of thing is just random."} -P03
\end{quote}

Perceptions of the interaction’s \textit{naturalness} also varied across participants, likely reflecting differences in user expectations and prior experience with artificial intelligence.
\begin{quote}    
    \textit{"People would like it, I think. How it could do things better used to be, I felt, at least I had the impression, that it sort of was programmed. It was sort of, you know, like pre-recorded track. It wasn't really engaging and following to what I'm saying to him."} -P01
\end{quote}
\begin{quote}  
    \textit{"The whole concept of talking to an AI thing, I thought, was really unnatural."} -P03
\end{quote}

Interestingly, this sentiment was not shared by all participants. Many perceived that the interaction was in fact natural and referenced feeling like the interaction was shared with another real person.

\begin{quote} 
    \textit{"It feels like you have someone to tap your back when you are excited. You need to be awake, man. Something like that... This one feels a bit organic."} -P31 (NCA)
\end{quote}
\begin{quote} 
    \textit{"It makes it more fun, I would say it was fun, like talking to someone, or it's pretty much like you're talking to him, and sitting next to you, so yeah, it's pretty much like that."} -P08
\end{quote}

\textbf{Individual differences in cognitive load}
Another key finding was that usability was shaped by participants’ expectations of how well the CA could regulate workload. Some perceived the conversation as too shallow to sustain alertness (\textit{underload}), while others experienced it as effortful or intrusive (\textit{overload}). Balancing these divergent needs emerged as a central challenge for CA design.

\begin{quote}
    Underload: \textit{"I think in the short term, it would be good, but I mean, how long can you maintain a surface-level conversation with the robot and still be engaged?"} -P12
\end{quote}
\begin{quote}
    Overload: \textit{"No one wants to carry a conversation while they’re driving, because that’s just increased effort."} -P12
\end{quote}

Overall, beyond the technological limitations already noted as affecting usability, participants’ perceptions of naturalness in real driving contexts were far from uniform. Some held very high expectations for natural conversational exchange, while others valued the agent more as meaningful companionship. In addition, participants' feedback emphasised that a CA should be able to adapt precisely to different users’ workload, striking a balance between underload and overload.

\section{Discussion}

\subsection{Emerging Archetypes of Conversational Agent Use}
\label{sec:archetypes}
A central contribution of this work is the demonstration that drivers recognised the CA as a useful safety feature. Across conditions and age groups, participants consistently described the CA as helping to wake drivers up. This baseline provides strong evidence that CAs are perceived as legitimate safety tools, not simply novel entertainment for L3 driving. Taken together, these findings extend prior simulator-based work \cite{large2016stimulating, Large2019-ou, Mahajan2021-rl} by demonstrating that conversational dialogue can function as an authentic safety tool in a real L3 vehicle with a live LLM-based agent. This may provide novel empirical evidence of how real-time HCI interactions are experienced when embedded within an actual driving system.

Our multi-stage analysis (interviews, in-car video, and KSS) revealed distinct user preference profiles, which we interpret as archetypes. These profiles emerged inductively from qualitative content analysis and align with established CA archetype frameworks (Service/Productivity, Entertainment, Companion) \cite{Lessio2020-xn}. As participants repeatedly indicated, the value of the CA depended on whether it aligned with their own orientation to the driving experience. Some drivers welcomed short safety-focused prompts, others expected entertainment or novelty to relieve monotony, and still others valued the CA as a social presence. These divergent perspectives point to the existence of distinct user archetypes, patterns of preference and orientation that shaped not only perceptions of usefulness but also future intention to use. This is crucial to support continued acceptance of a CA system as mismatched user-agent needs affect intention-to-use \cite{Venkatesh2008-ep}.

Although our priority was to design a safety agent whose job was strictly to discuss the immediate driving environment, users explained they would simply turn off a device that continued to engage them in discussions that were not of interest/priority for them. Thus, accommodating these diverse user-needs by presenting \textit{personas} can be seen as key to promoting continued engagement with CAs in automated driving.

Beyond simulator-based findings in this domain, our on-road results show that a brief, natural exchange can function as a safety intervention without resorting to intrusive alarms, a design approach participants frequently described as “bringing [them] back to alertness.” This strengthens the claim that LLM-based conversational dialogue can operate as a gentle but effective attention management strategy in authentic L3 conditions, rather than as novel NDRT entertainment only. 




\subsubsection*{\textbf{Identification of Archetypes}}
The heterogeneity in responses can be understood through the Technology Acceptance Model 3 (TAM3) \cite{Venkatesh2008-ep}. While perceived usefulness was consistently acknowledged, it was insufficient to guarantee intention to use unless participants also experienced the CA as \textit{job relevant}. For safety-first drivers, the system’s narrow focus on road-related prompts matched their priorities: \textit{“To talk about the environment or about the road or about speed or the weather, something which you need to be noting anyway”} (P104). For these users, usefulness and relevance were tightly coupled, supporting adoption.

By contrast, other participants experienced a mismatch between the agent’s safety-first framing and their own needs. They tend to prioritise hedonistic needs. Entertainment-seeking drivers valued stimulation over narrow functionality: \textit{“It could be enjoyable… if the automated system can introduce different things, rather than just speaking. You can play music or something, or there is a visual screen… if you can, like, change from time to time the experience… that would be great as well”} (P25). Socially oriented drivers, meanwhile, emphasised companionship and empathy: \textit{“It was good. It was really helpful. I thought that, yeah, someone else is here to help me… it was feeling so good”} (P39). For these groups, safety benefits were recognised but did not align with their self-defined goals, leading to reluctance despite positive effects on alertness.

This mismatch is critical. Without job relevance, a system risks being perceived as intrusive or irritating even when drivers accept its potential benefits. In the context of TAM3, the absence of alignment between agent type and user orientation threatens future intention to use. Thus, safeguarding adoption requires recognising and addressing variation in user archetypes. This contrasts our design of a safety-priority unit which was hypothesised to appeal to the most amount of participants.

Here, our analysis draws on the work of \citet{Nielsen2019-fx}, emphasising the identification of patterns of user difference before fictionalising characters. Through thematic analysis of interviews, triangulated with behavioural and KSS data, we derived three archetypes: \textbf{Safety-First}, \textbf{Entertainment-Seeking}, and \textbf{Social-Connection Oriented} (see~\autoref{tab:user-archetypes}). To further conceptualise their implications, we positioned them against the design archetypes articulated by \citet{Lessio2020-xn}: Service/Productivity, Entertainment, and Companion. This alignment underscores that not all drivers benefit from a single CA type. Instead, each archetype corresponds to a distinct design orientation with its own criteria for acceptance.
Simultaneously, the data shows that traits and roles are not exclusive; real users (and real agents) can exhibit overlapping characteristics, which argues for designs that acknowledge fluidity rather than enforce singular types. Indeed, users displayed shifting needs and expectations throughout the drive.

\begin{table}[t]
\caption{Archetypes of Conversational Agent Users}
\label{tab:user-archetypes}
\resizebox{\columnwidth}{!}{%
\begin{tabular}{@{}ll@{}}
\toprule
\multicolumn{2}{c}{\textbf{Safety-First Users}} \\ \midrule
\textbf{Trait profile} &
  \begin{tabular}[c]{@{}l@{}}High monotony tolerance; very high safety concern; \\ high alertness; negligible hedonic motivation; very low social need.\end{tabular} \\ \cmidrule(l){2-2} 
\textbf{Motivation} &
  \begin{tabular}[c]{@{}l@{}}To remain alert and maintain vigilance during automated driving; \\ any interaction must clearly serve a safety-related purpose.\end{tabular} \\ \cmidrule(l){2-2} 
\textbf{Acceptance Interpretation} &
  \begin{tabular}[c]{@{}l@{}}Perceived usefulness is the strongest determinant of intention to use. \\ Continued engagement depends on whether the conversational agent \\ clearly contributes to reduced fatigue and improved readiness.\end{tabular} \\ \cmidrule(l){2-2} 
\textbf{Illustrative quotes} &
  \begin{tabular}[c]{@{}l@{}}"To talk about the environment or about the road or about speed or the \\ weather, something which you need to be noting anyway. The problem is \\ people with passengers in the car, they often talk about things that are totally \\ unrelated." -P104\end{tabular} \\ \midrule
\multicolumn{2}{c}{\textbf{Entertainment-Seekers}} \\ \midrule
\textbf{Trait profile} &
  \begin{tabular}[c]{@{}l@{}}Very high hedonic motivation; moderate social need; moderate safety concern; \\ low alertness; low monotony tolerance.\end{tabular} \\ \cmidrule(l){2-2} 
\textbf{Motivation} &
  \begin{tabular}[c]{@{}l@{}}To combat boredom and make driving more stimulating through novelty, \\ humour, or game-like interactions.\end{tabular} \\ \cmidrule(l){2-2} 
\textbf{Acceptance Interpretation} &
  \begin{tabular}[c]{@{}l@{}}Perceived enjoyment dominates intention to use. Safety benefits are accepted \\ as secondary outcomes of staying mentally stimulated.\end{tabular} \\ \cmidrule(l){2-2} 
\textbf{Illustrative quotes} &
  \begin{tabular}[c]{@{}l@{}}"I think it's enjoyable, but not most. But it's enjoyable. It could be enjoyable… \\ if the automated system can introduce different things, rather than just speaking. \\ You can play music or something, or there is a visual screen… if you can, like, \\ change from time to time the experience given by the automated system, that \\ would be great as well." -P25\end{tabular} \\ \midrule
\multicolumn{2}{c}{\textbf{Social-Connection Oriented Users}} \\ \midrule
\textbf{Trait profile} &
  \begin{tabular}[c]{@{}l@{}}Very high social need; moderate hedonic value; moderate safety concern; \\ moderate alertness; moderate monotony tolerance.\end{tabular} \\ \cmidrule(l){2-2} 
\textbf{Motivation} &
  \begin{tabular}[c]{@{}l@{}}To experience companionship and empathic interaction during \\ monotonous driving; relational quality is central.\end{tabular} \\ \cmidrule(l){2-2} 
\textbf{Acceptance Interpretation} &
  \begin{tabular}[c]{@{}l@{}}Intention to use is driven by a combination of usefulness, enjoyment, and \\ social influence. The system must feel engaging and socially rewarding \\ while still contributing to alertness.\end{tabular} \\ \cmidrule(l){2-2} 
\textbf{Illustrative quotes} &
  \begin{tabular}[c]{@{}l@{}}"It was good. It was really helpful. I thought that, yeah, someone else \\ is here to help me… it was feeling so good." -P39\end{tabular} \\ \bottomrule
\end{tabular}%
}
\end{table}

\subsection{Linking Driver Archetypes to Conversational Agent Design Archetypes}

\citet{Lessio2020-xn} propose five CA design archetypes: \textit{Service, Companion, Entertainment, Care,} and \textit{Productivity}. We compared our three driver archetypes with this framework:

\begin{itemize}
    \item \textbf{Safety-First Users $\rightarrow$ Service \& Productive Archetype}

 Safety-first drivers align most closely with the \textit{Service} archetype, which is designed to provide clear, reliable, and functional support. Like service-oriented CAs (e.g., Alexa in its productivity focused mode), these drivers value task alignment, precision, and transparency over relational or hedonic qualities.
    \item \textbf{Entertainment-Seekers $\rightarrow$ Entertainment Archetype}

 Entertainment-driven drivers map directly to the \textit{Entertainment} archetype. These CAs emphasise play, humour, and novelty, sustaining engagement by offering stimulation. The fit is particularly strong with Nielsen’s finding that perceived enjoyment predicts continued use in this group \cite{Nielsen2019-fx}.
    \item \textbf{Social-Connection Oriented Users $\rightarrow$ Companion Archetype}

 Socially motivated drivers correspond to the \textit{Companion} archetype. Companion CAs, such as Replika\footnote{https://replika.com} or XiaoIce\footnote{https://www.xiaoice.com}, prioritise empathy, memory, and continuity of relationship. For these drivers, relational depth is the primary design quality that sustains acceptance and engagement.
\end{itemize}

By systematically identifying archetypes and mapping them to established CA design archetypes, we show that not all drivers benefit from CAs in the same way. Instead, \textit{safety-first, entertainment, and social-connection oriented users} align with distinct CA archetypes—Service, Entertainment, and Companion, respectively. This mapping provides both empirical grounding (through participant accounts) and future design heuristics (by connecting to a recognised archetype framework). Crucially, tailoring CA design to these archetypes with \textit{agent personas} may increase \textit{intention to use} as theorised by the acceptance model \cite{Venkatesh2008-ep}, thereby ensuring not only engagement but also improved safety outcomes in semi-automated driving contexts.

\subsection{Effectively Design CAs for Attention Management}
\subsubsection*{\textbf{CA for richer capabilities}}
Returning to our findings, driver responses, subjective reports, and interviews consistently suggest that CAs can help reduce passive fatigue in an authentic L3 driving context. KSS ratings showed opposite trends before and after interaction with the CA compared to the control group. This aligns with prior simulator-based results \cite{large2016stimulating}.

Participants also pointed out performance limitations. Similar constraints were noted in recent simulator studies, which found that low-complexity dialogue was most effective \cite{Huang2024-qe}. While this outcome seems intuitive, prior work rarely engaged with a broad set of users in situ to discuss deeper design trade-offs. Our study also introduced environment-related prompts that supported driver alertness, reducing the attention gap between automated and manual driving \cite{Lee2019-gf}. Although this paper did not include eye-tracking data, video analysis showed more searching behaviour after CA use, whereas control participants showed no similar change.

Positive behaviours were also observed. Some drivers laughed, joked, or otherwise displayed enjoyment when interacting with the CA. These reactions point to the agent’s potential to raise arousal through playfulness. Cognitive activities were also noted, such as thoughtful expressions, suggesting that the CA could increase cognitive load in monotonous driving contexts. A few even gestured towards the CA, unintentionally bringing their hands closer to the wheel and increasing physical readiness. Compared to monotonous NDRTs previously used to mitigate passive fatigue \cite{Naujoks2018-su}, the CA offered richer capabilities.

\subsubsection*{\textbf{Safety cues: explicit or implicit?}}
The overarching goal remains to provide effective guidance: preventing underload while supporting driver alertness to facilitate safe L3 driving, and doing so with smooth transitions that preserve user experience \cite{Mehrotra2022-sj}. The challenge is how to embed safety functions into CA interactions. User engagement, as discussed in Section~\ref{sec:archetypes}, is a prerequisite: drivers should be willing to use the system and not find it irritating. But should safety cues be made explicit or remain implicit? Our findings highlight three considerations:
\begin{enumerate}
    \item \textbf{Mode awareness.} Some participants argued that talking while driving was unsafe, even conflicting with road safety campaigns. This shows the importance of situating CAs in the right contexts (i.e. right modes) and building accurate mental models of their use.
    \item \textbf{Trust in automation.} Several participants, already in a heightened state of alertness due to distrust of AVs, found the CA distracting rather than helpful. Adaptive timing of interventions, possibly informed by a driver monitoring system, is therefore essential.
    \item \textbf{Conversation topics.} A few participants expressed interest in deeper interaction, such as playing puzzle games to “get the brain working.” While promising, overly complex dialogue risks cognitive overload and degraded performance \cite{McWilliams2021-lz}. Future LLM-based CAs should therefore integrate control strategies to balance the user requests for in-depth conversation. Designers can manipulate 
    (i) gate conversational depth by workload cues; 
    (ii) offer archetype switching via quick controls (e.g., “Safety-only”, “Safety+Entertainment”, “Safety+Social”); and 
    (iii) anchor enrichment (games, music, local info, small talk) to explicit user preference and driver state.
\end{enumerate}

In summary, our study surfaces design considerations critical for balancing safety and user experience in authentic automated driving contexts. Yet open questions remain. For example, how long can conversation effects persist before drivers revert to fatigue? What alternative or escalated strategies are effective when the agent fails to meet user expectations? These questions extend beyond the scope of this paper but represent important avenues for future research.

\section{Limitations}
This study presents early-stage insights based on a prototype AV platform and an LLM-based CA deployed in a dynamic field setting. While the on-road environment adds ecological validity, both the vehicle and the agent were prototypes. The AV operated on a rule- and map-based platform rather than production-level autonomy, and the agent occasionally experienced network-related delays due to the setting. The agent latency fluctuated, which may have provided an inconsistent experience between users. However, this is comparable to standard online voice-based LLM applications, and most participants still rated interactions as natural and responsive. 

Participants undertook a single 55-minute session, meaning long-term adoption, novelty effects, or habituation remain unexplored. Indeed, some participants suggested that their perceptions may change given time and exposure. While the sample was diverse in age, targeted recruitment of older and younger drivers to facilitate usability exploration across the age range may affect fatigue ratings. Specifically, older drivers may be less prone to fatigue manipulations in AVs \cite{Arefnezhad2022-yb}. Additionally, although no age-related differences in perceptions of usability emerged during qualitative analysis, between-subjects hypothesis testing may reveal distinct differences. 

While participant allocation favoured the CA condition, all participants experienced the same vehicle, route, environmental, and procedural conditions. None of the participants reported prior experience with L3 during screening or interviews, and such exposure was therefore not considered a confounding factor. Individual differences in communication style or familiarity with CAs were treated as natural variation relevant to understanding diverse user responses rather than as variables to be experimentally controlled.

We also acknowledge that anthropomorphic and personality cues may shape user trust, engagement, and perceptions of agency \cite{Ruijten2018-ir, Park2024-yh}. The agent’s gendered name (“Zoe”) may likewise have contributed to user perceptions \cite{Roesler2022-tv}, and future work should consider how naming conventions and gender presentation interact with these social dynamics.

Beyond this consensus, our analysis revealed three preference orientations (Safety-First, Social-Connection, and Entertainment-Seeking). We deliberately describe these as archetypes rather than fixed categories. Participants’ needs and orientations were not static, and individuals often displayed features of more than one profile or shifted between them during the drive. This fluidity is not a limitation of the finding but an opportunity for design. LLM-powered CAs are uniquely suited to accommodate dynamic shifts in user preferences through adaptive dialogue. Covering the three primary orientations observed here, therefore, provides a practical foundation for designing agents that can meet the majority of user needs while remaining sensitive to contextual changes.

Together, these limitations point to the need for deeper longitudinal, cross-cultural, and adaptive investigations. At the same time, they highlight the value of our study as a proof of concept: demonstrating that even under constrained conditions, conversational interaction can be deployed authentically in a real AV, yielding insights into how users experience, accept, and desire such agents.

\section{Conclusion}

This study demonstrates the potential of CAs to mitigate passive fatigue in conditional automated driving. Rather than relying on direct safety alerts, the approach embedded safety goals into natural, environment-related dialogue. In a real-world test-track study, users consistently identified the agent as a legitimate, prospective safety tool. Users described how brief conversational exchanges woke them up and re-engaged their attention with the driving environment, helping manage passive fatigue symptoms. Perceptions of the entertainment value of the agent were mixed. Overall, these findings provide strong evidence that LLM-powered conversational dialogue can serve as a credible intervention to support safety in L3 driving contexts.

Our analysis also revealed three predominant user preference profiles, Safety-First, Entertainment-Seeking, and Social-Connection oriented drivers. Aligning these orientations with established CA design archetypes \cite{Lessio2020-xn} highlights a central HCI contribution. Acceptance is not determined by usefulness alone but by the fit between agent behaviour and drivers’ own goals. For some, safety cues were paramount; for others, stimulation or companionship mattered most. Designing for conditional automation, therefore, requires agents that can flexibly accommodate diverse orientations without compromising safety. These insights provide preliminary design heuristics for future passive fatigue management CA development.  

Methodologically, the work advances simulator and Wizard-of-Oz approaches \cite{large2016stimulating, Large2019-ou} to study live LLM-based agents in L3 AVs. The integration of behavioural video, sleepiness ratings, and qualitative interviews offers a reusable model for evaluating conversational interaction in safety-critical domains. This approach demonstrates how early-stage prototypes can still yield valuable insights into user acceptance, experience, and usability evaluations.

Together, these contributions position conversational agents as future adaptive HCI interventions that bridge the gap between passive supervision and active readiness. By systematically linking empirical archetypes to design archetype frameworks and acceptance models, the paper provides conceptual and practical guidance for designing usable and desirable CAs in future AVs. This alignment offers practical guidance for tailoring agent personalities and interaction strategies to diverse driver needs, while recognising that user orientations may shift dynamically over time. These insights can be generalised to other safety-critical supervisory contexts, from aviation to rail and control rooms, where cognitive underload continues to threaten vigilance. By offering proof-of-concept and design guidance for fatigue-intervention conversational agents, we show how small talk can yield big impacts for the future of human–machine teaming.

\begin{acks}
This research was undertaken under the QUT Industry Chair in Empathic Machines, in collaboration with and co-funded by industry partner Seeing Machines Ltd. Through participation in the National Industry PhD Program (NIPP), it has also been co-funded by the Department of Education. The Australian Research Council also provided financial support for this project through the Australian Research Council Industrial Transformation Training Centre for Automated Vehicles in Rural and Remote Regions (IC230100001).

We thank the Queensland Department of Transport and Main Roads for providing access to the highly automated prototype vehicle Zoe2. We also thank the entire research team for their contributions, in particular the safety drivers who ensured operational safety throughout the study. We further acknowledge Sean Baek from the Faculty of Science for his support in software development, the research assistants who supported data collection, and the staff at the RACQ Mobility Centre for their support and understanding during the on-track data collection.
\end{acks}

\bibliographystyle{ACM-Reference-Format}
\bibliography{coversational-agent-base.bib}

@String{Computing = "Computing" }

@String{Springer = "Springer-Verlag" }

@techreport{SAEJ3016_202104,
    author = {{SAE International}},
    title = {Taxonomy and Definitions for Terms Related to Driving Automation Systems for On-Road Motor Vehicles},
    institution = {SAE International},
    year = {2021}
}

@ARTICLE{Merat2019-ph,
  title    = "The “Out-of-the-Loop” concept in automated driving: proposed
              definition, measures and implications",
  author   = "Merat, Natasha and Seppelt, Bobbie and Louw, Tyron and Engström,
              Johan and Lee, John D and Johansson, Emma and Green, Charles A and
              Katazaki, Satoshi and Monk, Chris and Itoh, Makoto and McGehee,
              Daniel and Sunda, Takashi and Unoura, Kiyozumi and Victor, Trent
              and Schieben, Anna and Keinath, Andreas",
  journal  = "Cogn. Technol. Work",
  volume   =  21,
  number   =  1,
  pages    = "87--98",
  month    =  feb,
  year     =  2019,
  url      = "https://doi.org/10.1007/s10111-018-0525-8"
}

@ARTICLE{Vogelpohl2019-ap,
  title    = "Asleep at the automated wheel-Sleepiness and fatigue during highly
              automated driving",
  author   = "Vogelpohl, Tobias and Kühn, Matthias and Hummel, Thomas and
              Vollrath, Mark",
  journal  = "Accid. Anal. Prev.",
  volume   =  126,
  pages    = "70--84",
  month    =  may,
  year     =  2019,
  url      = "http://dx.doi.org/10.1016/j.aap.2018.03.013",
  language = "en"
}

@INCOLLECTION{Hancock2017-fv,
  title     = "Conceptualizing and Defining Fatigue",
  author    = "Hancock, P A and Desmond, Paula A and Matthews, Gerald",
  booktitle = "The Handbook of Operator Fatigue",
  address = "London",
  publisher = "CRC Press",
  edition   = "1st Edition",
  pages     = "63--73",
  month     =  nov,
  year      =  2017,
  url       = "http://doi.org/10.1201/9781315557366-4"
}

@ARTICLE{Mahajan2021-rl,
  title    = "Exploring the benefits of conversing with a digital voice
              assistant during automated driving: A parametric duration model of
              takeover time",
  author   = "Mahajan, Kirti and Large, David R and Burnett, Gary and Velaga,
              Nagendra R",
  journal  = "Transp. Res. Part F Traffic Psychol. Behav.",
  volume   =  80,
  pages    = "104--126",
  month    =  jul,
  year     =  2021,
  url      = "https://doi.org/10.1016/j.trf.2021.03.012"
}

@inproceedings{large2016stimulating,
  title     = {Stimulating Conversation: Engaging Drivers in Natural Language Interactions with an Autonomous Digital Driving Assistant to Counteract Passive Task-Related Fatigue},
  author    = {Large, David R. and Burnett, Gary and Antrobus, Vicki and Skrypchuk, Lee},
  booktitle = {5th International Conference on Driver Distraction and Inattention (DDI2017)},
  pages     = {1--17},
  year      = {2019},
  publisher = {IFSTTAR},
  url       = {https://nottingham-repository.worktribe.com/output/2034939Task-Related_Fatigue},
  address = {Paris, France}
}

@ARTICLE{May2009-vc,
  title     = "Driver fatigue: The importance of identifying causal factors of
               fatigue when considering detection and countermeasure
               technologies",
  author    = "May, Jennifer F and Baldwin, Carryl L",
  journal   = "Transp. Res. Part F Traffic Psychol. Behav.",
  publisher = "Elsevier BV",
  volume    =  12,
  number    =  3,
  pages     = "218--224",
  month     =  may,
  year      =  2009,
  url       = "http://dx.doi.org/10.1016/j.trf.2008.11.005",
  language  = "en"
}

@INCOLLECTION{Kuckartz2019-ae,
  title     = "Qualitative text analysis: A systematic approach",
  author    = "Kuckartz, Udo",
  booktitle = "ICME-13 Monographs",
  publisher = "Springer International Publishing",
  address   = "Cham",
  pages     = "181--197",
  year      =  2019,
  url       = "http://doi.org/10.1007/978-3-030-15636-7_8",
  language  = "en"
}

@ARTICLE{McWilliams2021-lz,
  title     = "Underload on the road: Measuring vigilance decrements during
               partially automated driving",
  author    = "McWilliams, Thomas and Ward, Nathan",
  journal   = "Front. Psychol.",
  publisher = "Frontiers Media SA",
  volume    =  12,
  pages     =  631364,
  month     =  apr,
  year      =  2021,
  url       = "http://doi.org/10.3389/fpsyg.2021.631364",
  language  = "en"
}

@INPROCEEDINGS{Brown2023-bg,
  title     = "The Halting problem: Video analysis of self-driving cars in
               traffic",
  author    = "Brown, Barry and Broth, Mathias and Vinkhuyzen, Erik",
  booktitle = "Proceedings of the 2023 CHI Conference on Human Factors in
               Computing Systems",
  publisher = "ACM",
  address   = "New York, NY, USA",
  pages     = "1--14",
  month     =  apr,
  year      =  2023,
  url       = "http://doi.org/10.1145/3544548.3581045",
  language  = "en"
}

@BOOK{Heath2010-xw,
  title     = "Video in Qualitative Research",
  author    = "Heath, Christian and Hindmarsh, Jon and Luff, Paul",
  publisher = "SAGE Publications",
  address   = "Thousand Oaks, CA",
  series    = "Introducing Qualitative Methods Series",
  month     =  feb,
  year      =  2010,
  url       = "http://doi.org/10.4135/9781526435385",
  language  = "en"
}

@ARTICLE{Endsley1995-ll,
  title     = "Toward a Theory of Situation Awareness in Dynamic Systems",
  author    = "Endsley, Mica R",
  journal   = "Hum. Factors",
  publisher = "SAGE Publications Inc",
  volume    =  37,
  number    =  1,
  pages     = "32--64",
  month     =  mar,
  year      =  1995,
  url       = "https://doi.org/10.1518/001872095779049543"
}

@inproceedings{Yu2025-nl,
  title     = "Leveraging {LLM}-based Conversational Agents to Combat Passive Fatigue in Conditional Automated Driving Contexts",
 booktitle = "Proceedings of the 17th International Conference on Automotive User Interfaces and Interactive Vehicular Applications",
  address = "Brisbane, AU",
  author    = "Yu, Yueteng and Cockram, Lewis and Pardo Gaytan, Jorge and Li,
               Xiaomeng and Rakotonirainy, Andry and Kuo, Jonny and Demmel,
               Sebastien and Lenné, Mike and Schroeter, Ronald",
  publisher = "Association for Computing Machinery (ACM)",
  month     =  aug,
  year      =  2025,
  pages = "1--5",
  url       = "https://doi.org/10.1145/3744335.3758493"
}

@INPROCEEDINGS{Huang2024-qe,
  title     = "Chatbot and fatigued driver: Exploring the use of {LLM}-based
               voice assistants for driving fatigue",
  author    = "Huang, Shaoshuai and Zhao, Xuandong and Wei, Dapeng and Song,
               Xinheng and Sun, Yuanbo",
  booktitle = "Extended Abstracts of the CHI Conference on Human Factors in
               Computing Systems",
  publisher = "ACM",
  address   = "New York, NY, USA",
  month     =  may,
  year      =  2024,
  number    =  74,
  pages     = "1--8",
  url       = "http://doi.org/10.1145/3613905.3651031",
  language  = "en"
}

@INPROCEEDINGS{Motion-Sadeghian_Borojeni2018-gh,
  title     = "Feel the Movement: Real Motion Influences Responses to Take-over
               Requests in Highly Automated Vehicles",
  author    = "Sadeghian Borojeni, Shadan and Boll, Susanne C J and Heuten,
               Wilko and Bülthoff, Heinrich H and Chuang, Lewis",
  booktitle = "Proceedings of the 2018 CHI Conference on Human Factors in
               Computing Systems",
  publisher = "Association for Computing Machinery",
  address   = "Montreal QC Canada",
  number    = "Paper 246",
  pages     = "1--13",
  series    = "CHI '18",
  month     =  apr,
  year      =  2018,
  url       = "https://doi.org/10.1145/3173574.3173820"
}

@ARTICLE{Capallera2023-ux,
  title     = "Human-Vehicle Interaction to Support Driver's Situation Awareness
               in Automated Vehicles: A Systematic Review",
  author    = "Capallera, Marine and Angelini, Leonardo and Meteier, Quentin and
               Khaled, Omar Abou and Mugellini, Elena",
  journal   = "IEEE Transactions on Intelligent Vehicles",
  publisher = "IEEE",
  volume    =  8,
  number    =  3,
  pages     = "2551--2567",
  month     =  Mar,
  year      =  2023,
  url       = "http://doi.org/10.1109/TIV.2022.3200826"
}

@ARTICLE{Bolchini2007-as,
  title     = "A data-oriented survey of context models",
  author    = "Bolchini, Cristiana and Curino, Carlo A and Quintarelli, Elisa
               and Schreiber, Fabio A and Tanca, Letizia",
  journal   = "SIGMOD Rec.",
  publisher = "Association for Computing Machinery (ACM)",
  volume    =  36,
  number    =  4,
  pages     = "19--26",
  month     =  dec,
  year      =  2007,
  url       = "http://doi.org/10.1145/1361348.1361353",
  language  = "en"
}

@INPROCEEDINGS{Schilit1994-dr,
  title     = "Context-aware computing applications",
  author    = "Schilit, B and Adams, N and Want, R",
  booktitle = "1994 First Workshop on Mobile Computing Systems and Applications",
  address = "Santa Cruz, CA",
  publisher = "IEEE",
  pages     = "85--90",
  month     =  dec,
  year      =  1994,
  url       = "http://doi.org/10.1109/WMCSA.1994.16"
}

@book{creswell2010sage,
author = {Tashakkori, Abbas. and Teddlie, Charles.},
address = {Los Angeles},
booktitle = {Sage handbook of mixed methods in social & behavioral research},
title = {Sage handbook of mixed methods in social \& behavioral research },
edition = {2nd ed.},
isbn = {9781412972666},
year = {2010},
language = {eng},
lccn = {2010006465},
publisher = {SAGE Publications},
}

@INPROCEEDINGS{Gang2018-kt,
  title     = "Don't Be Alarmed: Sonifying Autonomous Vehicle Perception to
               Increase Situation Awareness",
  author    = "Gang, Nick and Sibi, Srinath and Michon, Romain and Mok, Brian
               and Chafe, Chris and Ju, Wendy",
  booktitle = "Proceedings of the 10th International Conference on Automotive
               User Interfaces and Interactive Vehicular Applications",
  publisher = "Association for Computing Machinery",
  address   = "New York, NY, USA",
  pages     = "237--246",
  series    = "AutomotiveUI '18",
  month     =  sep,
  year      =  2018,
  url       = "https://doi.org/10.1145/3239060.3265636"
}

@INPROCEEDINGS{Large2019-ou,
  title     = "It's small talk, jim, but not as we know it: engendering trust
               through human-agent conversation in an autonomous, self-driving
               car",
  author    = "Large, David R and Clark, Leigh and Burnett, Gary and Harrington,
               Kyle and Luton, Jacob and Thomas, Peter and Bennett, Pete",
  booktitle = "Proceedings of the 1st International Conference on Conversational
               User Interfaces",
  publisher = "ACM",
  address   = "New York, NY, USA",
  month     =  aug,
  year      =  2019,
  pages     = "1--7",
  number = 22,
  url       = "http://doi.org/10.1145/3342775.3342789",
  language  = "en"
}

@INCOLLECTION{Weinbeer2019-gy,
  title     = "Automated driving: The potential of non-driving-related tasks to
               manage driver drowsiness",
  author    = "Weinbeer, Veronika and Muhr, Tobias and Bengler, Klaus",
  booktitle = "Advances in Intelligent Systems and Computing",
  publisher = "Springer International Publishing",
  address   = "Cham",
  pages     = "179--188",
  series    = "Advances in intelligent systems and computing",
  year      =  2019,
  url       = "http://doi.org/10.1007/978-3-319-96074-6_19",
  language  = "en"
}

@ARTICLE{Loew2024-ew,
  title     = "How to counteract driver fatigue during conditional automated
               driving—A systematic review",
  author    = "Loew, Alexandra and Kurpiers, Christina and Götze, Martin and
               Nitsche, Sven and Bengler, Klaus",
  journal   = "Future Transportation",
  publisher = "MDPI AG",
  volume    =  4,
  number    =  1,
  pages     = "283--298",
  month     =  mar,
  year      =  2024,
  url       = "http://doi.org/10.3390/futuretransp4010015",
  language  = "en"
}

@ARTICLE{Naujoks2018-su,
  title     = "From partial and high automation to manual driving: Relationship
               between non-driving related tasks, drowsiness and take-over
               performance",
  author    = "Naujoks, Frederik and Höfling, Simon and Purucker, Christian and
               Zeeb, Kathrin",
  journal   = "Accid. Anal. Prev.",
  publisher = "Elsevier BV",
  volume    =  121,
  pages     = "28--42",
  month     =  dec,
  year      =  2018,
  url       = "http://doi.org/10.1016/j.aap.2018.08.018",
  language  = "en"
}

@ARTICLE{Stampf2022-xm,
  title     = "Towards implicit interaction in highly automated vehicles - A systematic literature review",
  author    = "Stampf, Annika and Colley, Mark and Rukzio, Enrico",
  journal   = "Proc. ACM Hum. Comput. Interact.",
  publisher = "Association for Computing Machinery (ACM)",
  volume    =  6,
  number    = "MHCI",
  pages     = "1--21",
  month     =  sep,
  year      =  2022,
  url       = "https://doi.org/10.1145/3546726",
  language  = "en"
}

@ARTICLE{Saxby2008-mu,
  title     = "Effect of active and passive fatigue on performance using a
               driving simulator",
  author    = "Saxby, Dyani J and Matthews, Gerald and Hitchcock, Edward M and
               Warm, Joel S and Funke, Gregory J and Gantzer, Thomas",
  journal   = "Proc. Hum. Factors Ergon. Soc. Annu. Meet.",
  publisher = "SAGE Publications",
  volume    =  52,
  number    =  21,
  pages     = "1751--1755",
  month     =  sep,
  year      =  2008,
  url       = "http://doi.org/10.1177/154193120805202113",
  language  = "en"
}

@ARTICLE{Bier2019-xr,
  title     = "Preventing the risks of monotony related fatigue while driving
               through gamification",
  author    = "Bier, Lukas and Emele, Michael and Gut, Kaja and Kulenovic, Jasna
               and Rzany, David and Peter, Max and Abendroth, Bettina",
  journal   = "Eur. Transp. Res. Rev.",
  publisher = "Springer Science and Business Media LLC",
  volume    =  11,
  number    =  1,
  pages     = "1--19",
  month     =  dec,
  year      =  2019,
  url       = "http://doi.org/10.1186/s12544-019-0382-4",
  language  = "en"
}

@INPROCEEDINGS{Lessio2020-xn,
  title     = "Toward design archetypes for conversational agent personality",
  author    = "Lessio, Nadine and Morris, Alexis",
  booktitle = "2020 IEEE International Conference on Systems, Man, and Cybernetics (SMC)",
  address = "Toronto, ON, Canada",
  publisher = "IEEE",
  pages     = "3221--3228",
  month     =  oct,
  year      =  2020,
  url       = "http://doi.org/10.1109/SMC42975.2020.9283254",
  language  = "en"
}

@INCOLLECTION{Lee2019-gf,
  title     = "Workload and attention management in automated vehicles",
  author    = "Lee, Joonbum and Venkatraman, Vindhya and Campbell, John L and
               Richard, Christian M",
  booktitle = "Human Performance in Automated and Autonomous Systems",
  publisher = "CRC Press",
  address   = "Boca Raton, FL : CRC Press/Taylor \& Francis Group, 2019.",
  edition   = "1st Edition",
  pages     = "213--230",
  month     = Sep,
  year      =  2019,
  url       = "https://doi.org/10.1201/9780429458330"
}

@INCOLLECTION{Baldwin2019-hu,
  title     = "Attention management in highly autonomous driving",
  author    = "Baldwin, Carryl L and McCandliss, Ian",
  booktitle = "Human Performance in Automated and Autonomous Systems",
  publisher = "CRC Press",
  address   = "Boca Raton, FL : CRC Press/Taylor \& Francis Group, 2019.",
  edition   = "1st Edition",
  pages     = "231--248",
  month     =  sep,
  year      =  2019,
  url       = "https://doi.org/10.1201/9780429458330"
}

@INPROCEEDINGS{van-Veen2017-we,
  title     = "Situation awareness in automated vehicles through proximal
               peripheral light signals",
  author    = "van Veen, Tom and Karjanto, Juffrizal and Terken, Jacques",
  booktitle = "Proceedings of the 9th International Conference on Automotive
               User Interfaces and Interactive Vehicular Applications",
  publisher = "Association for Computing Machinery",
  address   = "Oldenburg Germany",
  month     = Sep,
  year      =  2017,
  url       = "https://doi.org/10.1145/3122986.3122993",
  language  = "en",
  pages = "287--292"
}

@ARTICLE{Nees2016-fg,
  title    = "Speech Auditory Alerts Promote Memory for Alerted Events in a
              Video-Simulated Self-Driving Car Ride",
  author   = "Nees, Michael A and Helbein, Benji and Porter, Anna",
  journal  = "Hum. Factors",
  volume   =  58,
  number   =  3,
  pages    = "416--426",
  month    = May,
  year     =  2016,
  url      = "https://doi.org/10.1177/0018720816629279",
  language = "en"
}

@INPROCEEDINGS{Khorsandi2023-kg,
  title     = "{FabriCar}: Enriching the user experience of in-car media
               interactions with ubiquitous vehicle interiors using {E}-textile
               sensors",
  author    = "Khorsandi, Pouya M and Jones, Lee and Davoodnia, Vandad and
               Lampen, Timothy J and Conrad, Aliya and Etemad, Ali and Nabil,
               Sara",
  booktitle = "Proceedings of the 2023 ACM Designing Interactive Systems
               Conference",
  publisher = "ACM",
  address   = "New York, NY, USA",
  month     =  Jul,
  year      =  2023,
  pages = "1438--1456",
  url       = "https://doi.org/10.1145/3563657.3595988",
  language  = "en"
}

@inproceedings{yu2025-nd,
    title = "Designing With Motion: Exploring Vestibular Cues as a Subtle Awareness Nudge Modality in Automated Vehicles",
    author = "Yueteng Yu and Xiaomeng Li and Sebastien Demmel and Sebastien Glaser and Jonny Kuo and Mike Lenn{\'e} and Ronald Schroeter",
    year = "2025",
    month = jul,
    day = "15",
    pages = "1--11",
    url = "https://doi.org/10.1145/3744333.3747824",
    language = "English",
    booktitle = "17th International Conference on Automotive User Interfaces and Interactive Vehicular Applications (AutomotiveUI '25)",
    publisher = "Association for Computing Machinery (ACM)",
    address = "United States",
}

@ARTICLE{Venkatesh2008-ep,
  title     = "Technology acceptance model 3 and a research agenda on
               interventions",
  author    = "Venkatesh, Viswanath and Bala, Hillol",
  journal   = "Decis. Sci.",
  publisher = "Wiley",
  volume    =  39,
  number    =  2,
  pages     = "273--315",
  month     =  may,
  year      =  2008,
  url       = " https://doi.org/10.1111/j.1540-5915.2008.00192.x",
  language  = "en"
}

@INPROCEEDINGS{Hutchins2017-ov,
  title     = "Technology acceptance model for safety critical autonomous
               transportation systems",
  author    = "Hutchins, Nathan and Hook, Loyd",
  booktitle = "2017 IEEE/AIAA 36th Digital Avionics Systems Conference (DASC)",
  address = "St. Petersburg, Florida, USA",
  publisher = "IEEE",
  pages     = "1--5",
  month     =  sep,
  year      =  2017,
  url       = "http://doi.org/10.1109/DASC.2017.8102010",
  language  = "en"
}

@ARTICLE{Chen2023-hz,
  title     = "Analysis of Australian public acceptance of fully automated
               vehicles by extending technology acceptance model",
  author    = "Chen, Yilun and Khalid Khan, Shah and Shiwakoti, Nirajan and
               Stasinopoulos, Peter and Aghabayk, Kayvan",
  journal   = "Case Stud. Transp. Policy",
  publisher = "Elsevier BV",
  volume    =  14,
  number    =  101072,
  pages     =  101072,
  month     =  dec,
  year      =  2023,
  url       = "http://doi.org/10.1016/j.cstp.2023.101072",
  language  = "en"
}

@ARTICLE{Mehrotra2022-sj,
  title     = "Human-Machine Interfaces and Vehicle Automation: A Review of the
               Literature",
  author    = "Mehrotra, Shashank and Wang, Meng and Wong, Nicholas and Parker,
               Jah'inaya and Roberts, Shannon C and Kim, Woon and Romo, Alicia
               and Horrey, William J",
  journal   = "Accid. Anal. Prev.",
  publisher = "aaafoundation.org",
  volume    =  109,
  pages     = "18--28",
  year      =  2022,
  url       = "https://doi.org/10.1016/j.trf.2024.08.014"
}

@BOOK{Nielsen2019-fx,
  title     = "Personas - user focused design",
  author    = "Nielsen, Lene",
  publisher = "Springer",
  address   = "London, England",
  edition   =  2,
  series    = "Human-Computer Interaction Series",
  month     =  mar,
  year      =  2019,
  url       = "http://doi.org/10.1007/978-1-4471-7427-1",
  language  = "en"
}

@ARTICLE{Steinberger2017,
  title     = "From road distraction to safe driving: Evaluating the effects of
               boredom and gamification on driving behaviour, physiological
               arousal, and subjective experience",
  author    = "Steinberger, Fabius and Schroeter, Ronald and Watling,
               Christopher N",
  journal   = "Comput. Human Behav.",
  publisher = "Elsevier BV",
  volume    =  75,
  pages     = "714--726",
  month     =  oct,
  year      =  2017,
  url       = "http://doi.org/10.1016/j.chb.2017.06.019"
}

@ARTICLE{Arefnezhad2022-yb,
  title     = "Effects of automation and fatigue on drivers from various age
               groups",
  author    = "Arefnezhad, Sadegh and Eichberger, Arno and Koglbauer, Ioana
               Victoria",
  journal   = "Safety (Basel)",
  publisher = "MDPI AG",
  volume    =  8,
  number    =  2,
  pages     =  30,
  month     =  apr,
  year      =  2022,
  url       = "http://doi.org/10.3390/safety8020030",
  language  = "en"
}

@ARTICLE{Plutchik1982-cw,
  title     = "A psychoevolutionary theory of emotions",
  author    = "Plutchik, Robert",
  journal   = "Soc. Sci. Inf. (Paris)",
  publisher = "SAGE Publications",
  volume    =  21,
  number    = "4-5",
  pages     = "529--553",
  month     =  jul,
  year      =  1982,
  url       = "http://doi.org/10.1177/053901882021004003",
  language  = "en"
}

@ARTICLE{Young2002,
  title     = "Malleable Attentional Resources Theory: A New Explanation for the Effects of Mental Underload on Performance",
  author    = "Young, Mark S and Stanton, Neville A",
  journal   = "Hum. Factors",
  publisher = "SAGE Publications",
  volume    =  44,
  number    =  3,
  pages     = "365--375",
  month     =  sep,
  year      =  2002,
  url       = "https://doi.org/10.1518/0018720024497709",
  language  = "en"
}

@ARTICLE{Larue2011-vp,
  title     = "Driving performance impairments due to hypovigilance on
               monotonous roads",
  author    = "Larue, Grégoire S and Rakotonirainy, Andry and Pettitt, Anthony N",
  journal   = "Accid. Anal. Prev.",
  publisher = "Elsevier BV",
  volume    =  43,
  number    =  6,
  pages     = "2037--2046",
  month     =  nov,
  year      =  2011,
  url       = "http://doi.org/10.1016/j.aap.2011.05.023",
  language  = "en"
}

@ARTICLE{McKerral2023,
    title = {Supervising the self-driving car: Situation awareness and fatigue during highly automated driving},
journal = {Accident Analysis \& Prevention},
volume = {187},
pages = {107068},
year = {2023},
issn = {0001-4575},
url = {https://doi.org/10.1016/j.aap.2023.107068},
author = {Angus McKerral and Kristen Pammer and Cassandra Gauld}
}

@inproceedings{Schroeter2016,
author = {Schroeter, Ronald and Steinberger, Fabius},
title = {Pok\'{e}mon DRIVE: towards increased situational awareness in semi-automated driving},
year = {2016},
isbn = {9781450346184},
publisher = {Association for Computing Machinery},
address = {New York, NY, USA},
url = {https://doi.org/10.1145/3010915.3010973},
booktitle = {Proceedings of the 28th Australian Conference on Computer-Human Interaction},
pages = {25–29},
numpages = {5},
location = {Launceston, Tasmania, Australia},
series = {OzCHI '16}
}

@ARTICLE{Ruijten2018-ir,
  title     = "Enhancing trust in autonomous vehicles through intelligent user
               interfaces that mimic human behavior",
  author    = "Ruijten, Peter A M and Terken, Jacques M B and Chandramouli,
               Sanjeev N",
  journal   = "Multimodal Technol. Interact.",
  publisher = "MDPI AG",
  volume    =  2,
  number    =  4,
  pages     =  62,
  month     =  sep,
  year      =  2018,
  url       = "http://doi.org/10.3390/mti2040062",
  language  = "en"
}

@ARTICLE{Park2024-yh,
  title     = "Effects of autonomous driving context and anthropomorphism of
               in-vehicle voice agents on intimacy, trust, and intention to use",
  author    = "Park, Donggun and Lee, Yushin and Kim, Yong Min",
  journal   = "Int. J. Hum. Comput. Interact.",
  publisher = "Informa UK Limited",
  volume    =  40,
  number    =  22,
  pages     = "7179--7192",
  month     =  nov,
  year      =  2024,
  url       = "http://doi.org/10.1080/10447318.2023.2262271",
  language  = "en"
}

@ARTICLE{Roesler2022-tv,
  title     = "Why context matters: The influence of application domain on
               preferred degree of anthropomorphism and gender attribution in
               human–robot interaction",
  author    = "Roesler, Eileen and Naendrup-Poell, Lara and Manzey, Dietrich and
               Onnasch, Linda",
  journal   = "Int. J. Soc. Robot.",
  publisher = "Springer Science and Business Media LLC",
  volume    =  14,
  number    =  5,
  pages     = "1155--1166",
  month     =  jul,
  year      =  2022,
  url       = "http://doi.org/10.1007/s12369-021-00860-z",
  language  = "en"
}

@ARTICLE{Mortezapour2025-ys,
  title     = "Large language models in the service of ergonomics education: a
               theoretical discussion on extending the ergonomics curriculum
               through prompt engineering",
  author    = "Mortezapour, Alireza",
  journal   = "Theor. Issues Ergon.",
  publisher = "Informa UK Limited",
  volume    =  0,
  number    =  0,
  pages     = "1--11",
  month     =  oct,
  year      =  2025,
  url       = "http://doi.org/10.1080/1463922X.2025.2568125",
  language  = "en"
}

@INPROCEEDINGS{Braun2019-oh,
  title     = "At your service: Designing voice assistant personalities to
               improve automotive user interfaces",
  author    = "Braun, Michael and Mainz, Anja and Chadowitz, Ronee and Pfleging,
               Bastian and Alt, Florian",
  pages     = "1--11",
  number     = "40",
  booktitle = "Proceedings of the 2019 CHI Conference on Human Factors in
               Computing Systems",
  publisher = "ACM",
  address   = "New York, NY, USA",
  month     =  may,
  year      =  2019,
  url       = "http://doi.org/10.1145/3290605.3300270",
  language  = "en"
}

\newpage
\appendix

\begin{table*}
\caption{Appendix: Participants' demographic table}
\label{table:p_1}
\begin{tabular}{@{}llrllll@{}}
\toprule
PID & Gender & \multicolumn{1}{l}{Age} & Driving Exp. & Driving Frequency & AV Exp. (L1+) & CA/NCA \\ \midrule
P01  & Male   & 43 & 5+ yrs  & 5-7 days per week              & Yes & CA  \\
P02  & Female & 73 & 5+ yrs  & At least once in the last year & No       & CA  \\
P03  & Female & 48 & 5+ yrs  & 3-4 days per week              & No       & CA  \\
P04  & Female & 40 & 5+ yrs  & 5-7 days per week              & No       & CA  \\
P05  & Male   & 31 & 5+ yrs  & 5-7 days per week              & Yes      & CA  \\
P06  & Male   & 72 & 5+ yrs  & 5-7 days per week              & No       & NCA \\
P07  & Female & 31 & 2-5 yrs & At least once every month      & Yes      & CA  \\
P08  & Female & 31 & 5+ yrs  & At least once every month      & No       & CA  \\
P12  & Male   & 38 & 5+ yrs  & 1-2 days per week              & Yes      & CA  \\
P16  & Female & 44 & 5+ yrs  & 3-4 days per week              & No       & CA  \\
P22  & Male   & 32 & 5+ yrs  & 1-2 days per week              & No       & CA  \\
P25  & Male   & 32 & 5+ yrs  & 3-4 days per week              & Yes      & CA  \\
P29  & Male   & 44 & 5+ yrs  & 3-4 days per week              & Yes      & NCA \\
P31  & Male   & 34 & 5+ yrs  & At least once in the last year & No       & NCA \\
P36  & Female & 32 & 5+ yrs  & 5-7 days per week              & No       & NCA \\
P39  & Female & 33 & 5+ yrs  & At least once every month      & No       & CA  \\
P43  & Male   & 77 & 5+ yrs  & 5-7 days per week              & No       & CA  \\
P49  & Female & 81 & 5+ yrs  & 5-7 days per week              & No       & NCA \\
P50  & Female & 81 & 5+ yrs  & 5-7 days per week              & No       & CA  \\
P62  & Male   & 69 & 5+ yrs  & 3-4 days per week              & Yes      & CA  \\
P64  & Male   & 67 & 5+ yrs  & 5-7 days per week              & No       & NCA \\
P65  & Male   & 85 & 5+ yrs  & 3-4 days per week              & Yes      & CA  \\
P79  & Male   & 73 & 5+ yrs  & 5-7 days per week              & No       & NCA \\
P80  & Female & 72 & 5+ yrs  & 1-2 days per week              & No       & NCA \\
P84  & Female & 82 & 5+ yrs  & 1-2 days per week              & No       & CA  \\
P85  & Male   & 74 & 5+ yrs  & 5-7 days per week              & No       & CA  \\
P88  & Female & 68 & 5+ yrs  & 5-7 days per week              & No       & NCA \\
P89  & Male   & 82 & 5+ yrs  & 5-7 days per week              & No       & CA  \\
P94  & Male   & 36 & 5+ yrs  & 3-4 days per week              & Yes      & NCA \\
P95  & Female & 67 & 5+ yrs  & 5-7 days per week              & No       & CA  \\
P99  & Male   & 71 & 5+ yrs  & 5-7 days per week              & No       & CA  \\
P100 & Female & 33 & 5+ yrs  & 5-7 days per week              & Yes      & NCA \\
P104 & Male   & 73 & 5+ yrs  & 3-4 days per week              & Yes      & CA  \\
P106 & Male   & 50 & 5+ yrs  & 5-7 days per week              & No       & NCA \\
P107 & Male   & 66 & 5+ yrs  & 5-7 days per week              & No       & NCA \\
P108 & Male   & 35 & 5+ yrs  & 5-7 days per week              & Yes      & NCA \\
P110 & Male   & 47 & 5+ yrs  & 5-7 days per week              & No       & NCA \\
P111 & Female & 77 & 5+ yrs  & 3-4 days per week              & No       & CA  \\
P112 & Male   & 66 & 5+ yrs  & 5-7 days per week              & No       & CA  \\
P113 & Male   & 80 & 5+ yrs  & 5-7 days per week              & No       & CA  \\ \bottomrule
\end{tabular}
\end{table*}

\end{document}